\documentclass[namedreferences]{solarphysics}
\title{arxiv paper} 
\usepackage[optionalrh,showbiblabels]{spr-sola-addons} 
\usepackage{graphicx}        
\usepackage{color}           

\usepackage{multirow}
\usepackage{array}
\usepackage{longtable}
\usepackage{pdflscape}
\usepackage{rotating}
\usepackage{enumerate}
\usepackage{caption}
\usepackage{booktabs}
\usepackage{geometry}
\usepackage{makecell}
\usepackage{ulem}

\newcommand{\adv}{    {\it Adv. Space Res.}} 
 
\newcommand{\aap}{    {\it Astron. Astrophys.}}

\newcommand{\apj}{    {\it Astrophys. J.}}
\newcommand{\apjl}{   {\it Astrophys. J. Lett.}}

\newcommand{\jastp}{  {\it J. Atmos. Solar-Terr. Phys.}} 
\newcommand{\jgr}{    {\it J. Geophys. Res.}}

\newcommand{\solphys}{{\it Solar Phys.}}

\newcommand{\sw}{    {\it Space Weather}}
\newcommand{\asr}{    {\it Adv. Space Res.}}
\newcommand{\swsc}{    {\it J. Space Weather Spac.}}
\newcommand{\jpcs}{    {\it J. Phys. Conf. Ser.}}
\newcommand{\anng}{    {\it Ann. Geophys.}}
\newcommand{\radmeas}{    {\it Radiat. Meas.}}
\chardef\us=`\_

\graphicspath{{./plots/}}

\newcolumntype{L}[1]{>{\raggedright\let\newline\\\arraybackslash\hspace{0pt}}m{#1}}
\newcolumntype{C}[1]{>{\centering\let\newline\\\arraybackslash\hspace{0pt}}m{#1}}
\newcolumntype{R}[1]{>{\raggedleft\let\newline\\\arraybackslash\hspace{0pt}}m{#1}}

\begin{document}

\begin{article}
\begin{opening}

\title{Solar Energetic Particle Forecasting Algorithms and Associated False Alarms\\ {\it Solar Physics}}

\author{B.~\surname{Swalwell}$^{1}$\sep
        S.~\surname{Dalla}$^{1}$\sep
        R.W.~\surname{Walsh}$^{1}$      
       }

\institute{$^{1}$ Jeremiah Horrocks Institute, University of Central Lancashire, Preston, PR1 2HE, UK.
                     email: \url{bswalwell@uclan.ac.uk}\\ 
             }

\runningauthor{Swalwell et al.}
\runningtitle{SEP Forecasting Algorithms and Associated False Alarms}

\begin{abstract}

Solar energetic particle (SEP) events are known to occur following solar flares and coronal mass ejections (CMEs). However some high-energy solar events do not result in SEPs being detected at Earth, and it is these types of event which may be termed ``false alarms''. 

We define two simple SEP forecasting algorithms based upon the occurrence of a magnetically well-connected CME with a speed in excess of 1500 km s\(^{-1}\) (``a fast CME'') or a well-connected X-class flare and analyse them with respect to historical data sets. We compare the parameters of those solar events which produced an enhancement of \(>\)40 MeV protons at Earth (``an SEP event'') and the false alarms.

We find that an SEP forecasting algorithm based solely upon the occurrence of a well-connected fast CME produces fewer false alarms (28.8\%) than one based solely upon a well-connected X-class flare (50.6\%). Both algorithms fail to forecast a relatively high percentage of SEP events (53.2\% and 50.6\% respectively).

Our analysis of the historical data sets shows that false alarm X-class flares were either not associated with any CME, or were associated with a CME slower than 500 km s\(^{-1}\); false alarm fast CMEs tended to be associated with flares of class less than M3. 

A better approach to forecasting would be an algorithm which takes as its base the occurrence of both CMEs and flares. We define a new forecasting algorithm which uses a combination of CME and flare parameters and show that the false alarm ratio is similar to that for the algorithm based upon fast CMEs (29.6\%), but the percentage of SEP events not forecast is reduced to 32.4\%.

Lists of the solar events which gave rise to \(>\)40 MeV protons and the false alarms have been derived and are made available to aid further study.

\end{abstract}
\end{opening}

\section{Introduction}
\label{sec:intro}

Solar Energetic Particles (SEPs) pose a significant radiation hazard to humans in space \citep{2004AdSpR..34.1347H} and in high-flying aircraft, particularly at high latitudes \citep{2005AdSpR..36.1627B}. They also may cause serious damage to satellites \citep{2000JGR...10510543F} and make high-frequency radio communications either difficult or impossible \citep{2005AnGeo..23..359H}. Accurate forecasting of the arrival of SEPs at locations near Earth is consequently vital.

SEPs are known to be energised by flares and coronal mass ejections (CMEs), processes which can take place within the same active region in close temporal association. Flares exhibiting high levels of energy emission in soft X-rays (SXR) and CMEs with high speeds have long been associated with a high likelihood of SEPs being detected at Earth (see \textit{e.g.} \citealp{2015SoPh..290..841D}). The bases for making such associations are studies of large numbers of events which are directed towards demonstrating the connection between flare and CME properties, and SEP events. These studies go on to look for correlations between event parameters and the proportion of associated solar event SEPs (\textit{e.g.} \citealp{2005SoPh..229..135B, 2012ApJ...756L..29C}). 

Whether SEPs are actually detected at Earth, however, may depend upon many different factors: the mechanism behind their acceleration, the energy and efficiency of that acceleration, the location of the acceleration site, whether or not the particles can escape into the interplanetary medium, and how they travel through it. 

It is not the case that SEPs are detected at Earth following all large flares and fast CMEs (\textit{e.g.} \citealp{2011SoPh..269..309K}). Solar events of this type, which might reasonably be expected to produce SEPs at Earth but which do not, may be termed ``false alarms''. Furthermore, some SEP events may follow smaller solar events, so that they are ``missed events'' for SEP forecasting algorithms based on intense flares and/or fast CMEs.

Many SEP forecasting tools base their prediction upon the observation of intense solar flares and/or radio bursts. For example, the Proton Prediction System proposed by \citep{1989AdSpR...9..281S} makes a forecast based upon flare intensity and position. It produces almost equal numbers of correct forecasts, false alarms and missed events \citep{2007JASTP..69...43K}.

The National Oceanic and Atmospheric Administration (NOAA) Space Weather Prediction Center (SPWC) uses a system named ``Protons" which is described by \citep{1999RadiatMeas..30..231}. The tool aims to forecast the arrival of SEPs near Earth following the detection of solar flares and radio bursts. \citealp{2008SpWea...6.1001B}, validated the system over a period between 1986 and 2004, and found that its false alarm rate was 55\%. The tool, however, is only used as a decision aid and the actual forecasts issued by SWPC have improved over time\footnote{http://www.swpc.noaa.gov/sites/default/files/images/u30/S1\ Proton\ Events.pdf}. \citealp{SWE:SWE20256}, combine SEP event statistics with real-time SEP observations to produce a forecast which changes dynamically.

\citealp{2009SpWea...7.4008L}, developed the Empirical model for Solar Proton Events Real Time Alert (ESPERTA)  method of SEP forecasting based upon flare size, flare location and evidence of particle acceleration and escape. Their emphasis was to maximise the time between the issue of an SEP event warning and the arrival of the particles, and their aim was to produce an automated forecasting tool with a view to issuing warnings of SEP events without human intervention. Whilst it is a significant improvement over the Protons tool, the false alarm rate was, nevertheless, between 30\% and 42\% \citep{0004-637X-838-1-59}. The FORcasting Solar Particle Events and Flares (FORSPEF) model, proposed by \citep{1742-6596-632-1-012075}, aims to make forecasts of both flares and SEPs. Its SEP forecasting algorithm is based upon a purely statistical approach, and has not yet been validated.

Other forecasting tools use different methods. It has also been shown that type II radio bursts at decametric\textendash hectometric (DH) wavelengths may be used to aid the forecasting of SEP events. \citealp{0004-637X-809-1-105}, have described a statistical relationship between DH type II radio bursts, the properties of the associated type III burst, and peak proton flux. During the period they analysed (2010 to 2013) they were able to make predictions of an SEP event with a false alarm rate of 22\%.

The Relativistic Electron Alert System for Exploration (REleASE) SEP forecasting tool \citep{SWE:SWE185} relies upon the fact that electrons will travel faster than protons, and will therefore arrive at 1 AU first. A forecast of expected proton flux is made based upon the real-time electron flux measurements.

Although the majority of currently operational data-based forecasting schemes make use of flare information, it is widely thought that the use of CME information would substantially improve algorithm performance. While from an operational point of view it is currently not trivial to obtain CME parameters in real time, it is important to compare the performance of flare-based \textit{versus} CME-based algorithms and determine whether a combination of flare and CME parameters within a forecasting tool may be beneficial.

Along with empirical forecasting algorithms which are based upon solar observations, several physics-based space weather forecasting tools have recently been developed (\textit{e.g.} the SOLar Particle ENgineering COde (SOLPENCO) \citep{2006AdSpR..37.1240A}, a solar wind simulation including a cone model of CMEs \citep{2010AdSpR..46....1L}, and the Solar Particle Radiation SWx (SPARX) model \citep{SWE:SWE20233}.

A catalogue of 314 SEP events and their parent solar events between 1984 and 2013 has been produced by \citealp{2016JSWSC...6A..42P}. It is expected that this database will provide a solid basis for the analysis of SEP events and the characteristics of their parent solar event. The catalogue does not, however, include information on those solar events which were false alarms. In order to improve SEP forecasting tools for space weather applications, an analysis of the characteristics of false alarm events should be carried out with a view to gaining an understanding of why SEPs were not observed.

Some statistical studies of SEP events and false alarms have been undertaken. Most take the same approach as \citeauthor{2016JSWSC...6A..42P} and \citeauthor{2009SpWea...7.4008L}, starting by considering the SEP events and then looking for the possible parent solar events. \citep{2014EP&S...66..104G} examined solar events during the early part of solar cycle 24, and considered why some which had very fast CMEs and large flares did not produce ground level enhancements of energetic particles as might have been expected. They suggested that poor latitudinal magnetic connectivity between the solar event and the Earth may have been an important factor. 

\citealp{2006ApJ...642.1222M}, examined a small number of CMEs with a speed greater than 900 km s\(^{-1}\) which had no radio signature of flare-related acceleration, and found that none produced conspicuous SEP events at Earth. Those authors argue, therefore, that a CME shock without an associated flare is not sufficient to produce SEPs.

\citealp{2007ApJ...665.1428W}, suggested that X-class flares not associated with any CME may occur closer to the magnetic centre of their source active region and may therefore be confined by overlying arcade magnetic fields. \citealp{2010SoPh..263..185K} investigated a small number of  these ``CME-less'' flares further, and argued that no SEP event might be expected following a flare which shows high peak emission in soft X-rays but which does not exhibit radio emission at decimetre and longer wavelengths.

Most of the large sample studies described above started by considering SEP events and then looked for possible parent solar events. In this paper we take a different approach. We start our analysis by considering solar events and determining whether an SEP event was measured at Earth a short time thereafter.
We focus on intense flares and fast CMES and define two possible forecasting algorithms, the first based solely on the occurrence of an intense flare and the second on that of a fast CME. The performance of the algorithms is quantified by evaluating them over historical datasets, and the characteristics of false alarms studied. In addition, missed events, \textit{i.e.} SEP events not forecast, are also identified and studied. Finally we discuss how a new algorithm which combines flare and CME properties may be introduced, resulting in better performance.

We provide lists of false alarms based upon the forecasting algorithms in order that they may form the basis of future studies and comparisons, together with a list of the solar events which produced \(>\)40 MeV protons. We analyse the properties of the false alarm events to determine whether reasons why they did not produce SEPs at Earth can be identified.
 
\section{False alarms and forecasting algorithms}
\label{sec:falsealarms}

A false alarm may simply be defined as ``a solar event which is predicted by a forecasting algorithm to produce SEPs at Earth but which fails to do so''. Specification of a forecasting algorithm and determination of its associated false alarms requires identification of:

\begin{enumerate}
\item The criteria and observational data sets by which a solar event is assigned a high likelihood of producing SEPs at Earth. Typically this will include identification of the type of solar event (\textit{e.g.} flare or CME) expected to produce SEPs, of a requirement on the intensity of the event (\textit{e.g.} a flare with peak SXR flux, \(f_\textrm{{\footnotesize sxr}}\), which exceeds a specified threshold intensity, \(f_\textrm{{\footnotesize thr}}\), or a CME with a speed \(v_\textrm{{\tiny CME}}\) which is faster than a threshold speed \(v_\textrm{{\footnotesize thr}}\)), of a positional requirement (\textit{e.g.} an event with a source region west of a given longitude), and possibly of other parameters.
 
\item The criteria by which it is determined that an SEP event has occurred or not. These will typically include specification of the instrument being used to measure particle flux intensity, of the species of particle examined and its energy range, and of the SEP intensity threshold, \(I_\textrm{{\footnotesize thr}}\), used to establish whether an SEP event was detected following a particular solar event. 

\item The method by which the solar event is associated with the SEP event.
\end{enumerate}

We discuss each of these requirements in Sections~\ref{sec:solareventparameters}, \ref{sec:sepeventparameters}, and \ref{sec:assocofevents} respectively.

\subsection{Solar event parameters}
\label{sec:solareventparameters}
As our source for CME data we have used the Co-ordinated Data Analysis Workshop (CDAW) CME catalogue\footnote{http://cdaw.gsfc.nasa.gov/CME\_list/index.html} \citep{2009EM&P..104..295G}. This catalogue is produced manually, CMEs being identified visually from images obtained by the C2 and C3 coronagraphs of the \textit{Large Angle and Spectrometric Coronagraph Experiment} (LASCO) \citep{1995SoPh..162..357B}) on board the \textit{Solar and Heliospheric Observatory} (SOHO) spacecraft. 

Information is published in the catalogue on various CME parameters including, \textit{inter alia}, the time it is first seen in the LASCO images, its width, and its position angle. CDAW publishes three values for the speed of CMEs in its catalogue, each calculated by different means: we use the first, the ``linear'' speed, which is obtained simply by fitting a straight line to the height-time measurements. Importantly, there is no information directly available from the catalogue as to whether the CME is Earth-directed, or from where on the solar disk it originated. This imposes serious limitations in analysing whether or not a particular CME is likely to produce SEPs at Earth.  

Solar flares are classified by their peak SXR emission as measured in the 1 - 8 \AA{} channel of the \textit{Geostationary Observational Environmental Satellites} (GOES) \citep{1975STIN...7628260G} \textit{X-ray Sensor} (XRS) instruments. Flares with a peak flux in this energy channel above 10\(^{-4}\) W m\(^{-2}\) are designated to be of class X; those with a peak flux between 10\(^{-5}\) and 10\(^{-4}\) W m\(^{-2}\) are of class M; classes C, B, and A are defined in a similar fashion. No single instrument has been in continuous operation since 1975, although the design has changed little over the years \citep{1994SoPh..154..275G}.

As our source for solar flare data we have used the GOES SXR Flare List which has been continuously maintained since 1975, and which may be downloaded from the website\footnote{http://www.helio-vo.eu/} of the \textit{Heliophysics Integrated Observatory} \citep{2011AdSpR..47.2235B}. 

In addition to reporting the maximum SXR intensity and the time of the start, peak and end of the flare, the GOES SXR Flare List also usually reports its heliographic co-ordinates. However there is a significant number of flares for which the list does not provide this information. In these cases we have used values for co-ordinates from the following sources:

\begin{enumerate}
\item Co-ordinates reported in the SolarSoft Latest Events Flares List (gevloc) (which may also be obtained through Helio). 

\item The reported co-ordinates of the active region (AR) from which the flare originated according to the GOES SXR flare list. 

\item Making our own estimate of co-ordinates by watching movies of 195 \r{A} images taken by \textbf{the} \textit{Extreme ultraviolet Imaging Telescope} (EIT) on board the SOHO spacecraft or of 195 \r{A} images taken by the \textit{Atmospheric Imaging Assembly} (AIA) on board the \textit{Solar Dynamics Observatory} (SDO).
\end{enumerate}
 
CMEs and solar flares, particularly high energy events, often occur within a short time of each other from the same solar active region. Making associations between these solar events is required so as to gain an understanding of the type of event which did, or did not produce SEPs at Earth: it also allows an estimate to be made of the site of origin of the CME from the reported heliographic coordinates of its associated flare.

We developed a method of making associations between CMEs and flares automatically which we set out in Appendix A. Whilst we are confident that the method produces correct associations in over 90\% of cases, to be sure we also viewed 195 \r{A} (obtained by the EIT on board SOHO) and 193 \r{A} (obtained by the AIA on board SDO) movies of each solar event. We confirmed the associations made by the automatic  method in 156 cases, changed them in six cases, and were unable to confirm the associations in a further 17 cases because EIT or AIA images were not available. 

\subsection{Location criterion for solar events}
\label{sec:west} 

\begin{figure}
\centerline{\includegraphics[width=.8\textwidth]{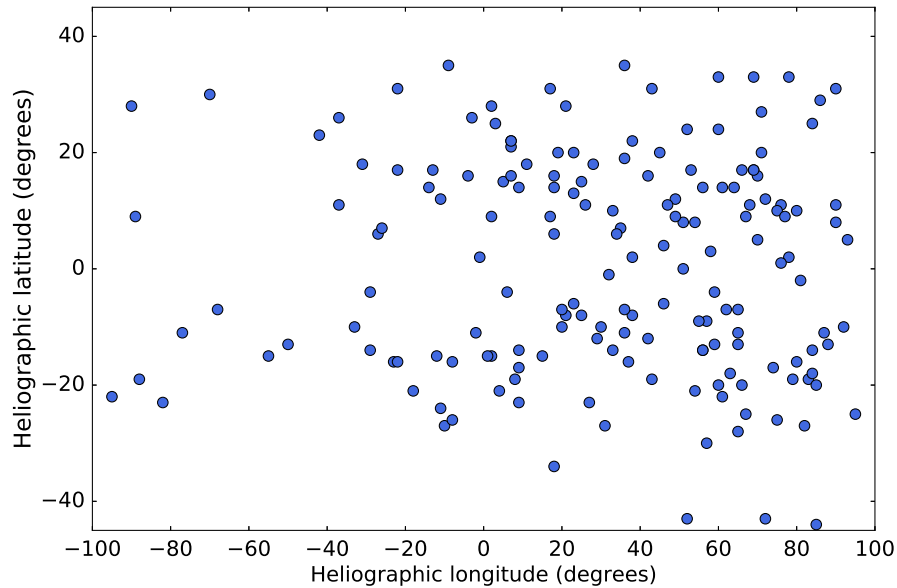}}
\caption{Heliographic longitude and latitude of solar events which produced an SEP event according to the criteria defined in Section~\ref{sec:sepeventparameters} between 1 April 1980 and 31 March 2013.}
\label{fig:SEPs}
\end{figure}

It is well known that solar events with origin in the west of the Sun as observed by an observer on Earth are more likely to produce SEPs than those originating in the east. Therefore it is common to introduce a positional criterion within SEP forecasting algorithms. Figure~\ref{fig:SEPs} shows the heliographic longitude of the 171 SEP-producing events between 1 April 1980 and 31 March 2013 for which we were able to determine coordinates. Of these, 86.5\% (148/171) had their origin in a solar event which occurred at a site west of E20, hence our choice of positional requirement in the forecasting algorithms.  We call solar events which have their origin west of E20 ``western events''. 

\subsection{The forecasting algorithms}
\label{sec:algorithms}

The two forecasting algorithms we investigate in this work are based upon the fact that that the more energetic the solar event, the greater the likelihood of that event producing SEPs at Earth, particularly if magnetically well-connected (\textit{e.g.} \citealp{2015SoPh..290..841D}). The algorithms are:

\begin{enumerate}[{A}.1]
\item A frontside CME with a reported speed of 1500 km s\(^{-1}\) or greater (a ``fast CME'') occurring west of E20 on the solar disk will result in an SEP event being detected at Earth.

\item An X-class flare occurring west of E20 on the solar disk will result in an SEP event being detected at Earth.
 
\end{enumerate}

We evaluate both the forecasting algorithms over the time range from 11 January 1996 until 31 March 2013 (``time range 1''); for algorithm A.2 we are also able to examine a longer period, between 1 April 1980 and 31 March 2013 (``time range 2''). In time range 1 there were 143 fast CMEs (according to our definition set out in A.1) reported by CDAW and 140 X-class flares. In time range 2 there were 403 X-class flares. 

Table~\ref{tab:coords} sets out the numbers of solar events which we have examined in this study. A number of solar events have had to be excluded from our analysis because of data gaps, the saturation of detectors or other cause, or because it was not possible to determine the heliographic co-ordinates.

\begin{table}
\caption{Numbers of solar events the subject of this study. Column 1 shows the time range over which data have been analysed, column 2 the type of solar event considered, column 3 the total number of solar events within the period investigated, column 4 the number of events for which we were able to determine coordinates (after removal of events discarded due to data gaps, saturation of detectors or other reasons) and column 5 the number of events which occurred west of E20.}
\begin{tabular}{l p{2cm}
>{\centering\arraybackslash}m{2cm} 
>{\centering\arraybackslash}m{2.2cm} 
>{\centering\arraybackslash}m{2.2cm}}
\toprule
Time range & Event type & Total number of events & 
Events for which coordinates were determined &
Analysed events west of E20 \\
\midrule
Time range 1 & Fast CMEs & 143 & 93 & 52 \\
\textbf{(Jan 1996 to Mar 2013)} & X-class flares & 140 & 139 & 79 \\
\midrule
Time range 2 &
\multirow{2}{*}{X-class flares} & 
\multirow{2}{*}{403} & 
\multirow{2}{*}{377} & 
\multirow{2}{*}{197} \\
\textbf{(Apr 1980 to Mar 2013)} & & & & \\
\bottomrule
\end{tabular}
\label{tab:coords}
\end{table}

\subsection{SEP event parameters}
\label{sec:sepeventparameters}
The definition of an SEP event will typically include specification of the instrument being used to measure particle flux, of the species of particle examined and its energy range, and of the SEP intensity threshold, \(I_\textrm{{\footnotesize thr}}\) used to establish whether an SEP event was detected following a particular solar event.

Particles accelerated by solar events include electrons, protons, and heavier ions, but we have chosen to analyse high energy (\(>\) 40 MeV) protons. The threshold considered is a little higher than the \(>\) 10 MeV threshold used by NOAA, making our event list less biased towards interplanetary shock-accelerated events. This choice also avoids proton enhancements caused by magnetospheric effects. 

\begin{figure}
\centerline{\includegraphics[width=0.6\textwidth]{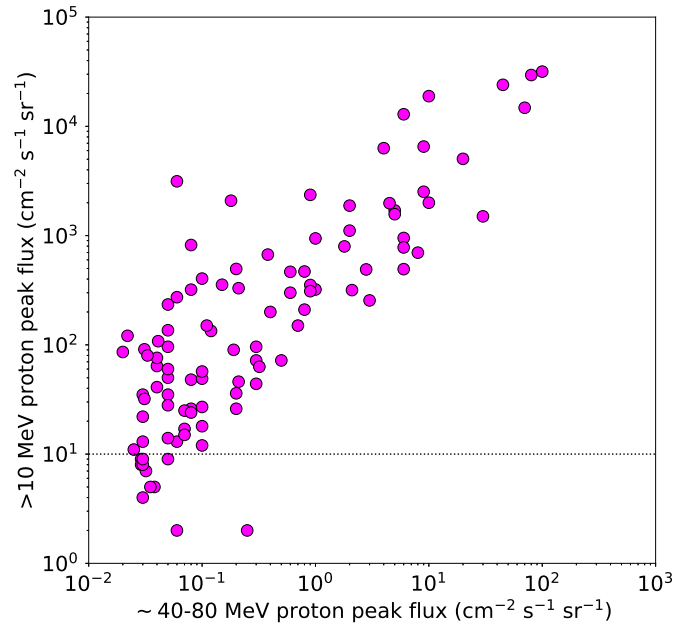}}
\caption{Peak flux of \(>\)10 MeV protons as reported by NOAA plotted against peak proton flux in the GOES \(\sim\)40-80 MeV energy channel in time range 1. The dotted horizontal line is at the NOAA threshold of 10 pfu.}
\label{fig:peak_flux_comparison}
\end{figure}

Because our threshold energy for protons is higher than that used by NOAA, we compared peak \(>\)40 MeV fluxes for our event sample with the peak \(>\)10 MeV fluxes for the same events. For each of our events a value for \(>\)10 MeV flux was obtained from the NOAA SEP list\footnote{ftp://ftp.swpc.noaa.gov/pub/indices/SPE.txt}. Eleven of the SEP events at \(>\)40 MeV did not reach the NOAA threshold of 10 pfu at \(>\)10 MeV, and for these we estimated peak flux by visual analysis of the plots of each event\footnote{Downloaded from https://solarmonitor.org/}. Figure~\ref{fig:peak_flux_comparison} is a plot of the peak flux of \(>\)10 MeV protons plotted against peak proton flux in the \(\sim\)40-80 MeV energy channel of the GOES EPS instruments for the SEP events in time range 1. The dotted horizontal line is at the NOAA threshold of 10 particles cm\(^{-2}\) s\(^{-1}\) sr\(^{-1}\) (pfu). The highest value for maximum peak flux at \(>\)40 MeV in time range 1 was approximately 100 pfu - the same event at \(>\)10 MeV produced 31700 pfu according to NOAA.

All instruments which detect proton intensities are subject to slight fluctuations, and not all of these can properly be said to be SEP events. The definition of intensity threshold, \(I_\textrm{\footnotesize {thr}}\), must be high enough so as to exclude the normal fluctuations in measurements, but low enough to ensure that rises which are genuinely due to solar events are included. We set \(I_\textrm{\footnotesize {thr}}\) to be a 2.5-fold increase in proton intensity over the quiet-time background level.

For this study we have used GOES SEP data because they allow us to study SEP events over a time period of more than 30 years. No one instrument has been in continuous operation during that time, and so we have had to use data from a number of different GOES satellites. Table~\ref{tab:instrs} sets out which spacecraft we have used and the energy channel considered to establish the occurrence of an SEP event. There are slight differences in the energy channels, particularly in the case of GOES 2, but we take the view that the differences are so small as to have a negligible effect upon our results. We downloaded data from the European Space Agency's \textit{Solar Energetic Particle Environment Monitor} (SEPEM) website \citep{2010cosp...38.4225C}\footnote{http://dev.sepem.oma.be/}. Data from 1 April 1987 onwards had been cleaned and intercalibrated by the SEPEM team; prior to that date we used their raw data.  

\begin{table}
\centering
\small
\caption{Instruments used to obtain data on proton intensity, the dates between which data from that instrument was used, and the energy channels which have been analysed. Column 1 gives the name of the spacecraft from which the data we have used was taken, column 2 the date from which we began to use those data and column 3 the date when we ceased using those data. Column 4 shows the range of proton energies measured by the instrument we have used, and column 5 whether the data was raw or had been cleaned by the SEPEM team.}
\begin{tabular}{l c c c c}
\toprule
Spacecraft & Start date & End date & Energy channel (MeV) & Raw data / Cleaned \\
\midrule

GOES 2 & 1 April 1980 & 31 December 1983 & 36.0 - 500.0 & Raw data \\
GOES 6 & 1 January 1984 & 31 March 1987 & 39.0 - 82.0 & Raw data \\
GOES 7 & 1 April 1987 & 28 February 1995 & 39.0 - 82.0 & Cleaned \\
GOES 8 & 1 March 1995 & 7 January 2003 & 40.0 - 80.0 & Cleaned \\
GOES 12 & 8 January 2003 & 31 December 2009 & 40.0 - 80.0 & Cleaned \\
GOES 11 & 1 January 2010 & 31 December 2010 & 40.0 - 80.0 & Cleaned \\ 
GOES 13 & 1 January 2011 & 31 March 2013 & 38.0 - 82.0 & Cleaned \\
\bottomrule
\end{tabular}
\label{tab:instrs}
\end{table}

It is not always easy to determine whether an SEP event had occurred if the instrument were still recording high-energy protons from a previous event. If it were the case that the intensity level had not returned to within 2.5 times the quiet-time background level by the time of the start of the solar event we were investigating, that solar event was disregarded - it could not be known whether or not that event produced SEPs at Earth. The only exceptions were those cases where there was a clear increase in proton intensity which could only be attributed to the solar event in question, in which case it was treated as an SEP event.

We determined that, during time range 2, there had been 221 flux enhancements in the GOES \(>\) 40 MeV proton channel which satisfied our definition of an SEP event. 

\subsection{Association of solar events and SEP events}
\label{sec:assocofevents}
A criterion for associating solar events and SEP enhancements is necessary. First we took the start time of the solar event. For CMEs not associated with a flare we used the time the CME was first reported in the CDAW catalogue; for CMEs which were associated with a flare and for all flares, we used the reported start time of the flare. 

We then searched searched the GOES proton data for a subsequent SEP event. In most cases the SEP enhancement began before another solar event was reported, in which case the association between the solar event and the SEP enhancement was made. In some instances, however, another solar event was reported before the SEP enhancement commenced. For these cases it was assumed that it was this new solar event which accelerated the particles unless that event was so close in time to the arrival of the SEPs (\(\sim\)20 minutes) that it was unlikely that the new event could have been the cause. None of our confirmed solar event - SEP association time differences was as short as 20 minutes.

A number of solar events had to be discarded because they coincided with gaps in SEP data, meaning that it could not be known whether or not they had produced an SEP enhancement. However, if there had been short outages (\(\sim\)3 hours), and there was no evidence of an SEP event either side of the outage, the solar event has been counted as a false alarm. 

We also associated solar events to all of the 221 proton events which we identified. In some cases the associated flare was of a class smaller than X and/or the associated CME was not a fast one according to our definition. Of these 221 events, we were not able to determine coordinates of the parent solar event for 50. The event was a western one in 148 of the remaining 171 cases.

\section{Identification of false alarms and evaluation of the forecasting algorithms}

We applied the forecasting algorithms described in Section~\ref{sec:algorithms} to the historical data sets we collected. We evaluated both algorithms over time range 1 (1996 to 2013) and in addition we evaluated algorithm A.2 over the longer time range 2 (1980 to 2013).

\subsection{Algorithms A.1 and A.2 over time range 1}
\label{sec:flaresandcmes}

\begin{figure}
\centerline{\includegraphics[width=0.6\textwidth]{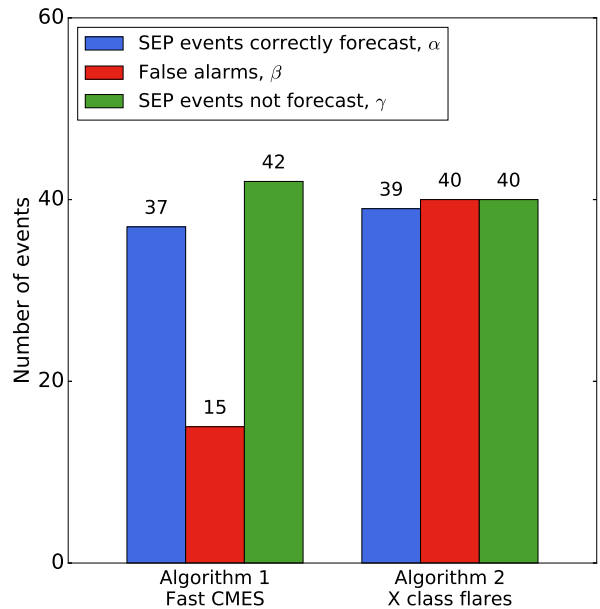}}
\caption{The numbers of correctly forecast SEP events, false alarms and SEP events which were not forecast for the two forecasting algorithms during time range 1.}
\label{fig:summary}
\end{figure}

Figure~\ref{fig:summary} shows the results of applying the two SEP forecasting algorithms to the data set for time range 1. The number of correctly forecast SEP events is shown by the blue bar and named \(\alpha\); the number of false alarms is represented by the red bar and named \(\beta\); and the number of SEP events which occurred but which were not forecast by the algorithm (the ``missed events'') is shown as the green bar and named \(\gamma\). There was a total of 107 SEP events in time range 1. Of the 86 SEP events for which we were able to determine the coordinates of the parent solar event, 91.9\% (79/86) were western events. 

Algorithm A.1 considers western fast CMEs. There were 52 such events during the period in question, and 71.2\% (37/52) produced SEPs at Earth. Thus the false alarm rate was 28.8\% (15/52) but the algorithm failed to forecast 53.2\% (42/79) of SEP events for which the parent solar event was a western one. Of all the SEP events for which coordinates could be determined, it missed 57.0\% (49/86).

Algorithm A.2 uses western X-class flares as the basis for the forecast. There were 79 such flares in time range 1, and 49.4\% (39/79) produced SEPs at Earth. The false alarm rate was therefore 50.6\% (40/79) and the algorithm failed to forecast 50.6\% (40/79) of SEP events for which the parent solar event was a western one. Of all the SEP events for which coordinates could be determined, it missed 54.7\% (47/86).

Appendix B provides the list of false alarms for the algorithm A.1, and Appendix C the false alarms for A.2 - the same lists are available electronically as supplementary material.

As well as reaching for an understanding of the underlying physical differences between those solar events which produced SEPs at Earth and the false alarms, we also look to measure the efficacy of the forecasting algorithms. A high percentage of correctly forecast SEP events (\(\alpha\)) coupled with a low number of false alarms (\(\beta\)) is desirable, but not at the expense of failing to forecast a large number of the SEP events which did occur (\(\gamma\)). In our evaluation we use two ratios: 

\begin{enumerate}
\item The ``false alarm ratio'' (FAR) gives the fraction of forecast events which actually did occur. It is defined as: 

\begin{equation}
 \textrm{FAR} = \frac{\beta}{\alpha + \beta}
\end{equation}

The FAR is sensitive to the number of false alarms, but takes no account of missed events. Possible scores range from 0 to 1, with the ``perfect'' score being 0.

\item The ``critical success index'' (CSI) is a measure of how well the forecast events correspond to the observed events. It is defined as 

\begin{equation}
 \textrm{CSI} = \frac{\alpha}{\alpha + \beta + \gamma}
\end{equation}

Possible scores range from 0 to 1, with the ``perfect'' score being 1. 
\end{enumerate}

\subsection{Forecasting algorithm A.1 - fast CMEs}
\label{sec:a1_results}

All the CMEs in our sample were from the front-side of the Sun and had an associated flare which was used to determine the coordinates. The FAR for algorithm A.1 is 0.29 and the CSI, not taking account of the missed eastern events, is 0.39. If the eastern events were to be included within the calculation for the CSI, its value would be reduced to 0.37. The evaluation scores for this algorithm over time range 1, and for algorithm A.2 over both time ranges, are summarised in Table~\ref{tab:scores}. It is not clear whether the high number of missed events is due to the fact that the measured velocity of the CME, \(v_\textrm{\tiny {CME}}\), is the plane-of-the-sky speed, whether in general the speeds measured by examination of coronagraph images are not sufficiently accurate, or whether more physics need to be included in the analysis.

{\centering
\begin{table}
\caption{A summary of the evaluation scores for the two forecasting algorithms: 
the ``false alarm ratio'' (FAR) and the ``critical success index'' (CSI) over 
time range 1. Algorithm A.2 is also evaluated over time range 2. Column 1 shows 
the forecasting algorithm being considered, column 2 the time range over which 
the analysis has been done, column 3 the false alarm ratio (FAR) for that 
algorithm, column 4 the critical success index (CSI) not taking into account the missed eastern events, and column 5 the CSI were these additional missed events to be included}.
\begin{tabular}{l c C{1cm} C{2.8cm} C{2.8cm}}
\toprule
Forecasting algorithm & Time range & FAR & CSI not including missed eastern events & CSI including missed eastern events \\
\midrule
A.1 (Fast CMEs) & 1 & 0.29 & 0.39 & 0.37 \\
A.2 (X-class flares) & 1 & 0.51 & 0.33 & 0.31 \\
A.2 (X-class flares) & 2 & 0.61 & 0.29 & 0.26 \\
\bottomrule
\end{tabular}
\label{tab:scores}
\end{table}}

\begin{figure}
\centerline{\includegraphics[width=.9\textwidth]{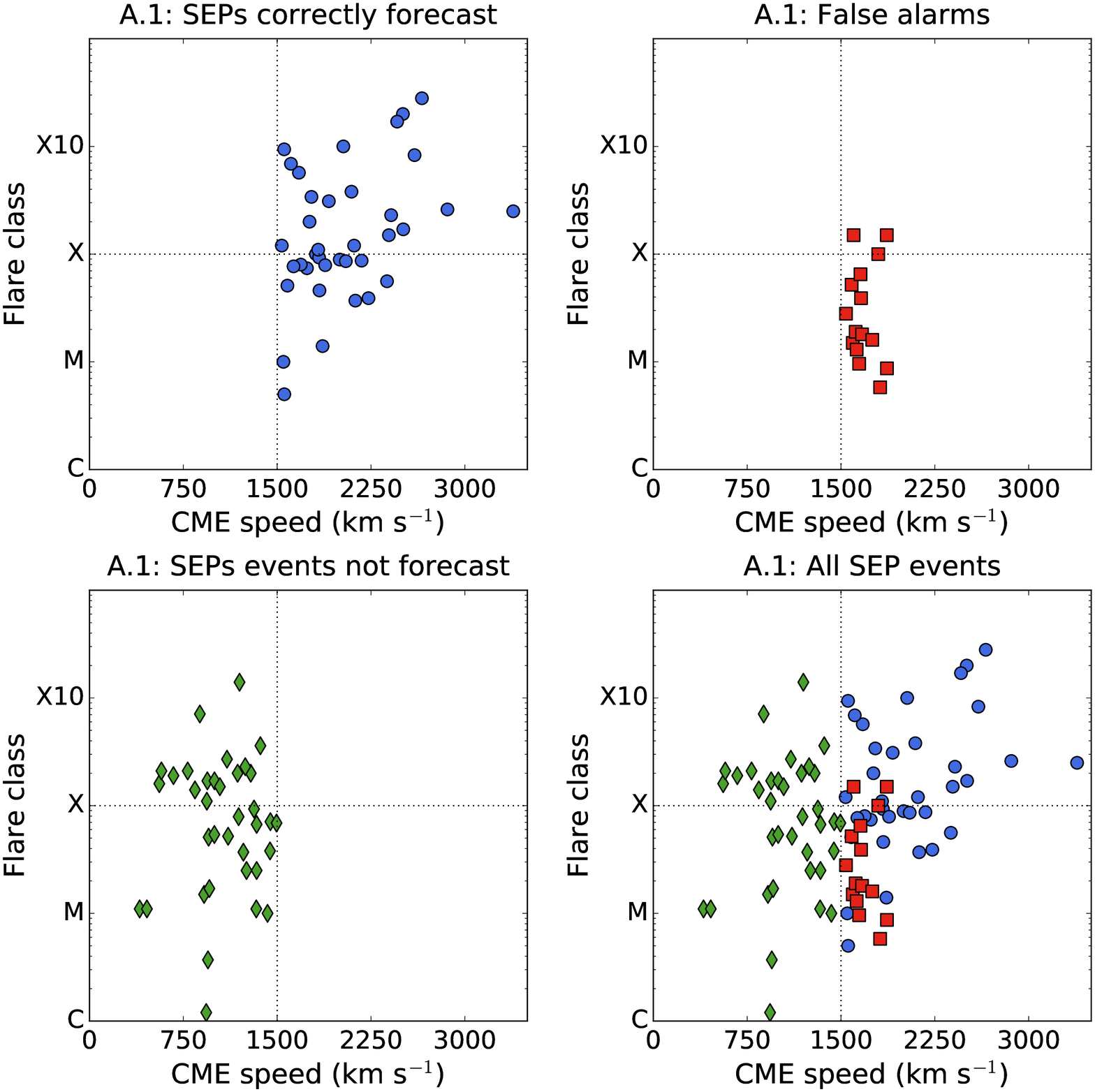}}
\caption{Flare class \textit{versus} associated CME speed for those 
solar events which produced SEPs $>$40 MeV at Earth in time range 1 (\textit{top left, blue circles}); for fast CMEs which were false alarms according to forecasting algorithm A.1 (\textit{top right, red squares}); for SEP events missed by algorithm A.1 (\textit{bottom left, green diamonds}); and for all events together (\textit{bottom right}).}
\label{fig:alg1_v_v_intensity}
\end{figure}

In Figure~\ref{fig:alg1_v_v_intensity} we plot peak SXR intensity of the CME's associated flare against its speed for those solar events in time range 1 which produced SEPs at Earth (top left, blue circles); for those events in the same period which were false alarms according to algorithm A.1 (top right, red squares); for SEP events missed by algorithm A.1 (bottom left, green diamonds); and for all events together (bottom right). Here one can see that many of the fast CME false alarms occur close to the threshold speed, \(v_\textrm{\footnotesize {thr}}\), and so increasing the threshold would reduce the number of false alarms, although it would also increase the number of missed events. A significant fraction of SEP events were associated with CMEs of reported speed much slower than 1500 km s\(^{-1}\). It is also clear that many of the false alarms have a flare intensity \(<\) M3. 

\begin{figure}
\centerline{\includegraphics[width=1\textwidth]{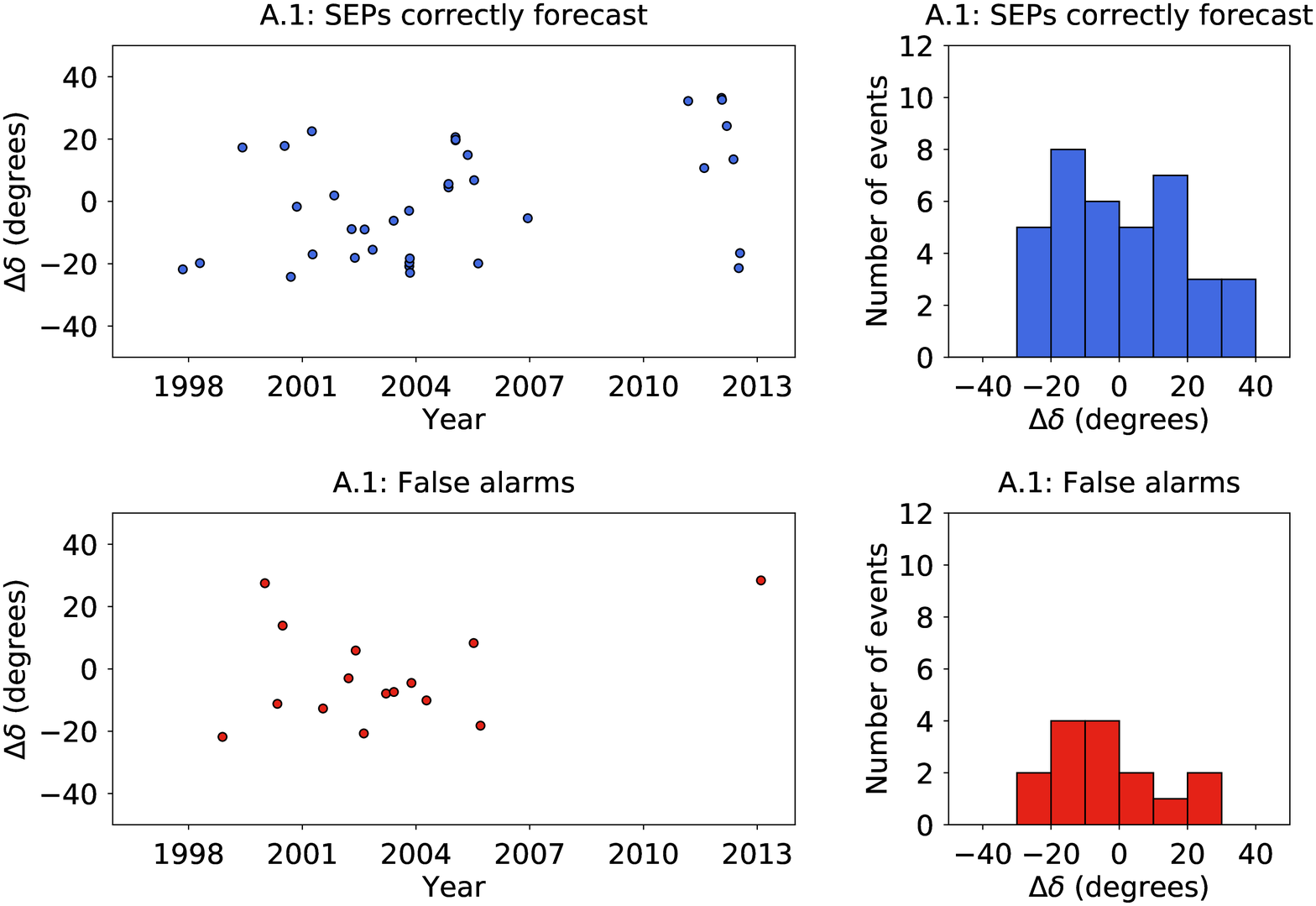}}
\caption{Plots of \(\Delta\delta\) against time for algorithm A.1, together with histograms of \(\Delta\delta\). The \textit{top plots} present the results for the solar events which were correctly forecast to produce SEPs at Earth (shown in \textit{blue}); the \textit{bottom plots} the false alarms (shown in \textit{red}).}
\label{fig:alg_1_hist_delta_delta_2_plots}
\end{figure}

\citealp{2014EP&S...66..104G} studied major solar eruptions during the first 62 months of solar cycle 24 and suggested that, among other things, the separation in latitude between the flare and the footpoint to Earth may be an important factor in determining whether high-energy particle events are detected. Therefore we define a parameter, \(\Delta\delta\), the difference between the latitude of the flare, \(\delta_\textrm{\footnotesize flare}\), and the latitude of the Earth's footpoint, \(\delta_\textrm{\footnotesize Earth}\), \textit{i.e.} the parameter \(\Delta\delta\) takes into account the inclination of Earth's orbit. In Figure~\ref{fig:alg_1_hist_delta_delta_2_plots} we plot \(\Delta\delta\) against time for Algorithm A.1, together with histograms for \(\Delta\delta\).  The events correctly forecast to produce SEPs are presented in the top plots (shown in blue), and the false alarms in the bottom plots (shown in red). For fast CMEs which had their origin within \(\pm\)10 degrees of the Earth's footpoint, 64.7\% (11/17) produced SEPs; for those which had their origin outside this range, 74.3\% (26/35) produced SEPs. Overall there does not appear to be a significant difference between the distribution in \(\Delta\delta\) for SEP events and false alarms. 

\begin{figure}
\centerline{\includegraphics[width=.9\textwidth]{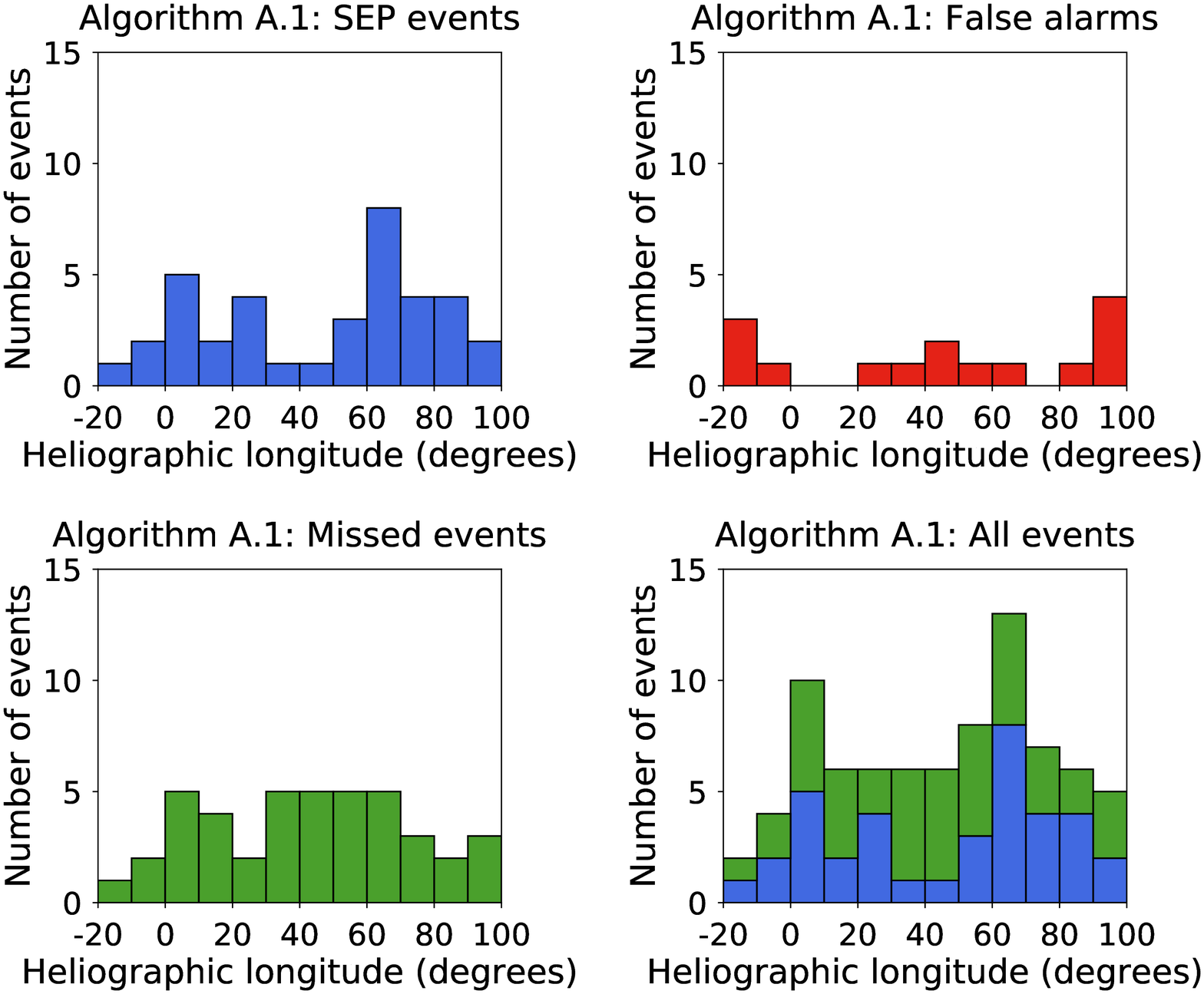}}
\caption{Histograms of the heliographic longitude of solar events in time range 1 of algorithm A.1 SEP events (\textit{top left}); of algorithm A.1 false alarms (\textit{top right}); of SEP events missed by algorithm A.1 (\textit{bottom left}); and of all SEP events (\textit{bottom right}).}
\label{fig:alg_1_hist_lon}
\end{figure}

Figure~\ref{fig:alg_1_hist_lon} shows histograms of the heliographic longitude of solar events in time range 1 correctly forecast by algorithm A.1 to produce an SEP event (top left), of algorithm A.1 false alarms (top right), of SEP events missed by algorithm A.1 (bottom left), and of all SEP events (bottom right). There is a peak of SEP-producing fast CMEs between W50 and W90. The false alarms for algorithm A.1 are relatively evenly distributed, as are the SEP events not forecast by A.1.

\begin{figure}
\centerline{\includegraphics[width=.85\textwidth]{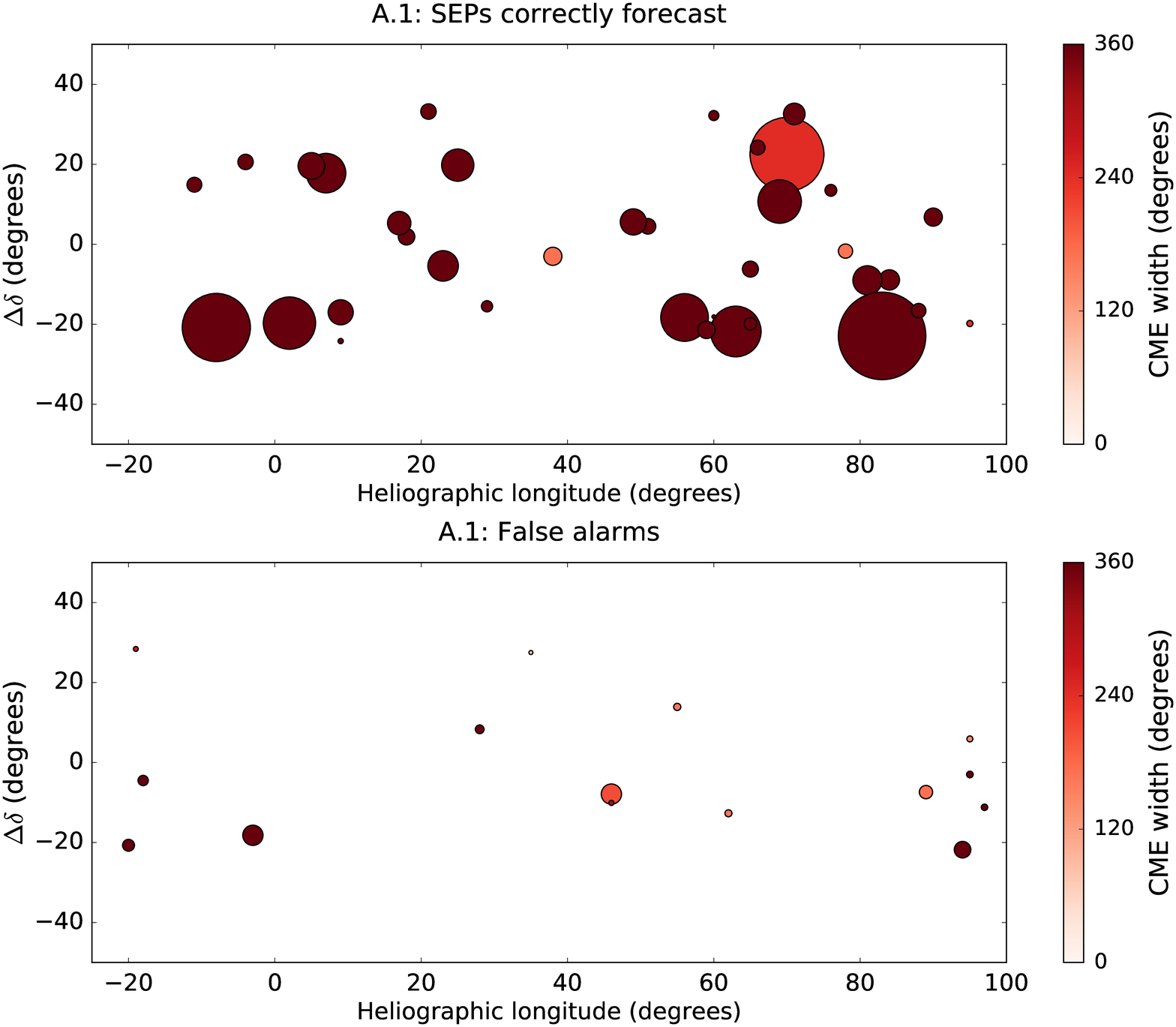}}
\caption{\(\Delta\delta\) \textit{versus} heliographic longitude for those western fast CMEs which produced SEPs at Earth in time range 1 (\textit{top plot}); and for those which were false alarms according to algorithm A.1 (\textit{bottom plot}). The size of the marker represents peak SXR intensity of the flare: for example, the point at S20W95 in the top plot was an M1.8 flare, whereas the point at S21E08 in the same plot was an X17.2 flare. The \textit{colour of the marker} represents CME width.}
\label{fig:alg1_lon_v_delta_delta_width_intensity}
\end{figure}

In Figure~\ref{fig:alg1_lon_v_delta_delta_width_intensity} we plot \(\Delta\delta\) against the longitude of the 37 western fast CMEs which produced an SEP event in time range 1. The size of the marker reflects the peak SXR intensity of the associated flare, and its colour is representative of the width of the CME. The bottom plot gives the same information, but for the false alarms according to algorithm A.1. It can be seen that, on average, the size of the markers in the middle plot is smaller than those in for the SEP-producing events. Thus, the peak SXR intensity of a fast CME's associated flare is relevant to the question as to whether SEPs will arrive at Earth. 

Also apparent from Figure~\ref{fig:alg1_lon_v_delta_delta_width_intensity} is that CME width is an important parameter. Of the 37 SEP-producing CMEs, 86.5\% (32/37) were reported to be haloes by the CDAW catalogue. By contrast, for the algorithm A.1 false alarms, only 46.7\% (7/15) were haloes. Therefore we find that halo CMEs are more likely to produce SEPs than non-haloes. This result is consistent with the findings of \citealp{2012JGRA..117.8108P} who found that solar events which had the highest probability of producing 10 MeV protons were full halo CMEs with a speed exceeding 1500 km s\(^{-1}\).

It should be noted that \citealp{2015ApJ...799L..29K} examined 62 halo CMEs (as reported by the CDAW catalogue) which occurred between 2010 and 2012 and were observed by three spacecraft separated in longitude by nearly 180\(^{o}\). They found that 42 were observed to be haloes by all three spacecraft. They concluded that a CME may appear to be a halo as a result of fast magnetosonic waves or shocks, and that apparent width does not represent an accurate measure of CME ejecta size.

\subsection{Forecasting algorithm A.2 - X-class flares}
\label{sec:a2_results}

Algorithm A.2 has an FAR of 0.51. Whilst it makes almost exactly the same number of correct forecasts as Algorithm A.1, the percentage of correct forecasts is lower. The proportion of missed SEP events is also relatively high, leading to a CSI of 0.33 without accounting for the missed eastern events, or of 0.31 if the missed eastern events were to be included. 

\begin{figure}
\centerline{\includegraphics[width=.8\textwidth]{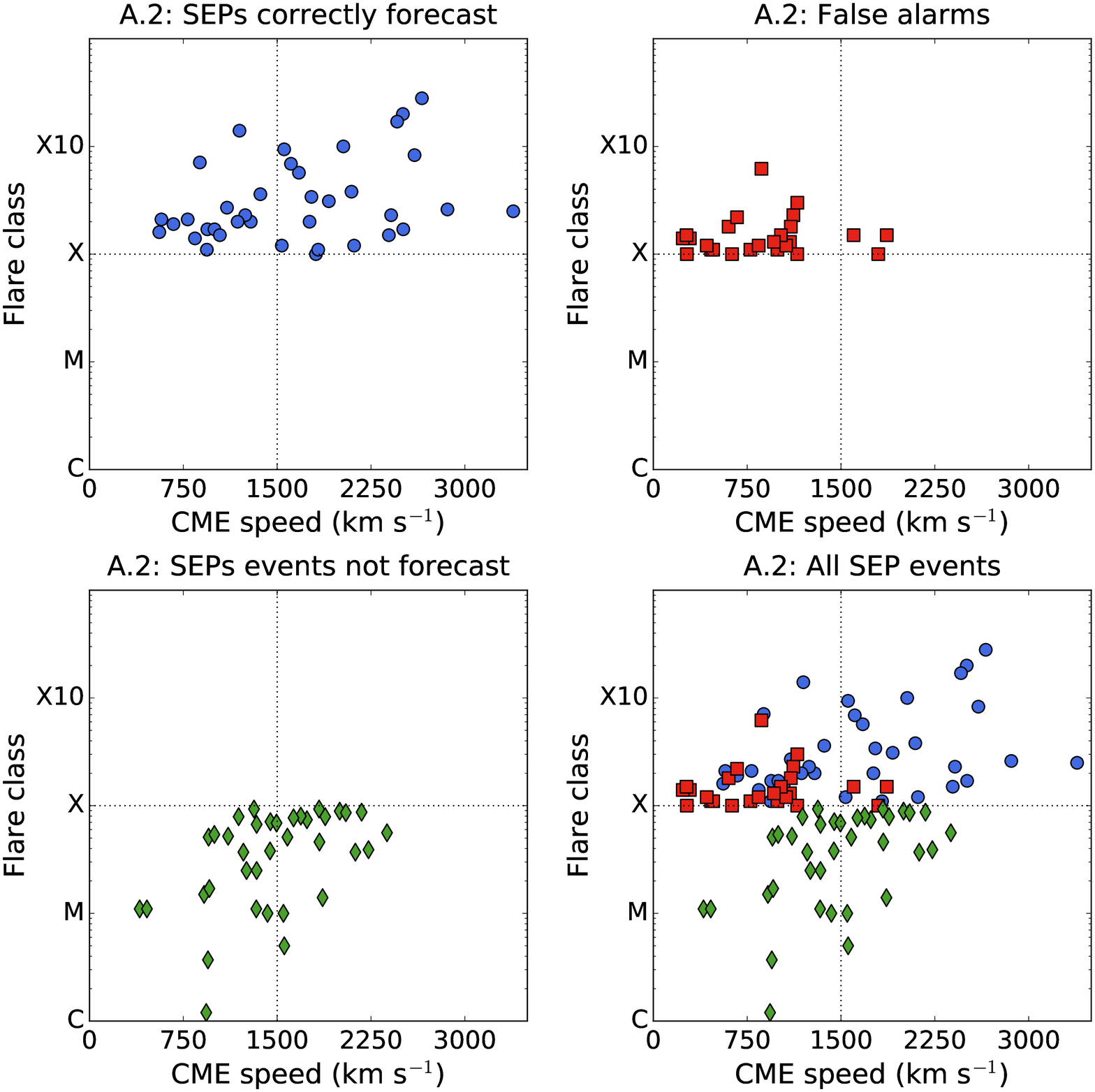}}
\caption{Flare class \textit{versus} associated CME speed for those solar events which produced SEPs $>$40 MeV at Earth in time range 1 (\textit{top left, blue circles}); for fast CMEs which were false alarms according to forecasting algorithm A.2 (\textit{top right, red squares}); for SEP events missed by algorithm A.2 (\textit{bottom left, green diamonds}); and for all events together (\textit{bottom right}).}
\label{fig:alg2_v_v_intensity}
\end{figure}

In Figure~\ref{fig:alg2_v_v_intensity} we plot SXR intensity for the solar flares above the threshold of A.2 against associated CME speed, and for SEP events missed by algorithm A.2 in the same format as in Figure~\ref{fig:alg1_v_v_intensity}. There is some symmetry with Figure~\ref{fig:alg1_v_v_intensity} in that many of the false alarms fall close to the chosen threshold. It should be noted that not all events above the A.2 threshold have an associated CME. Of the 122 X-class flares which occurred in time range 1 (and which did not coincide with a LASCO data gap), 14.8\% (18/122) had no associated CME. However the percentage of A.2 false alarms which did not coincide with a LASCO data gap and which did not have an associated CME is 26.5\% (9/34).

\begin{figure}
\centerline{\includegraphics[width=.9\textwidth]{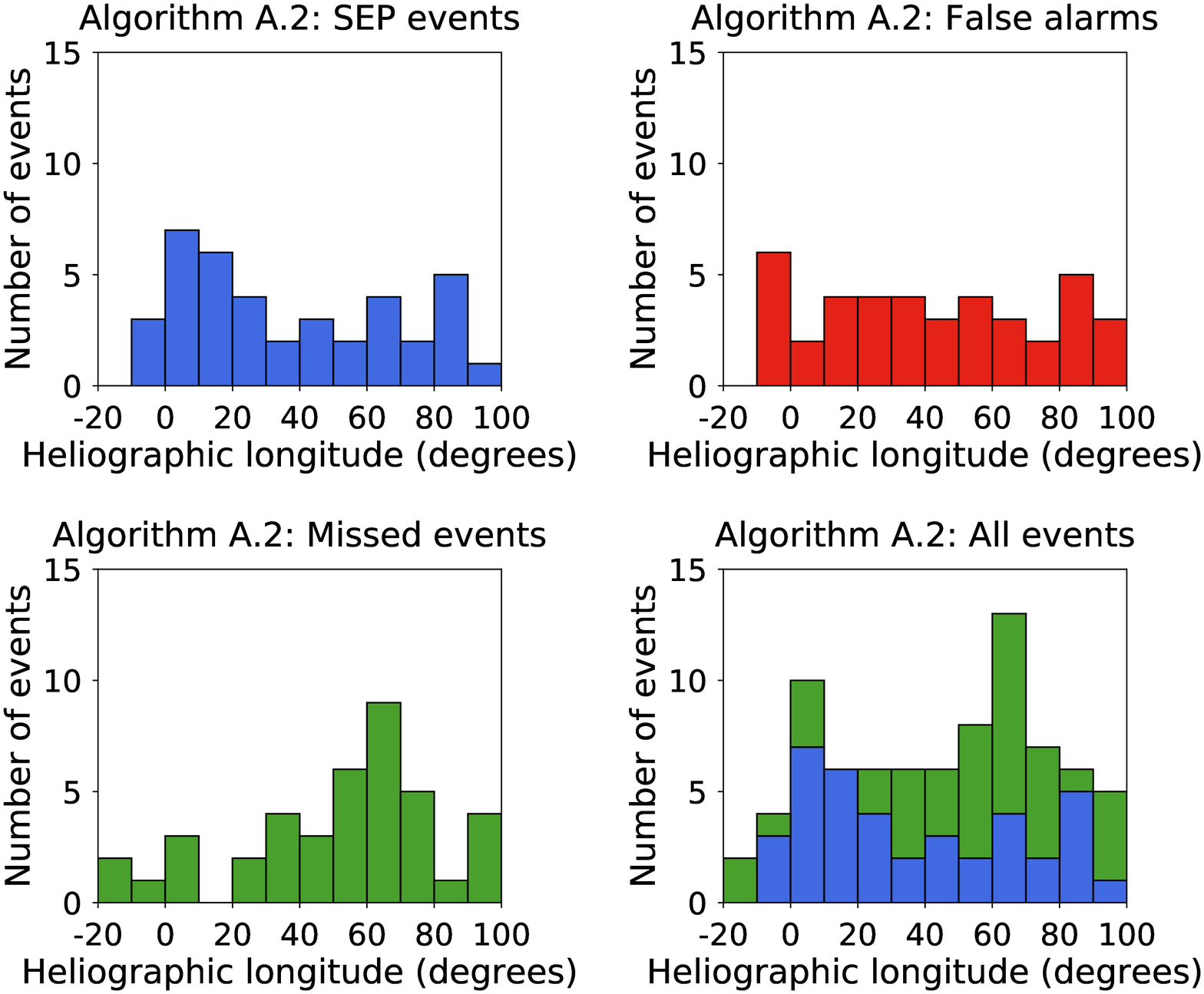}}
\caption{Histograms of the heliographic longitude of solar events in time range 1 of algorithm A.2 SEP events (\textit{top left}); of algorithm A.2 false alarms (\textit{top right}); of SEP events missed by algorithm A.2 (\textit{bottom left}); and of all SEP events (\textit{bottom right}).}
\label{fig:alg_2_hist_lon}
\end{figure}

In Figure~\ref{fig:alg_2_hist_lon} we show histograms of the heliographic longitude of solar events in time range 1 for algorithm A.2 in the same format as Figure~\ref{fig:alg_1_hist_lon}. There appears to be no significant difference in the longitudinal distribution of western X-class flares which produced an SEP event and those which were false alarms, but in this case the SEP events which were not forecast by algorithm A.2 do have a clear peak between W20 and W80. 

\begin{figure}
\centerline{\includegraphics[width=.85\textwidth]{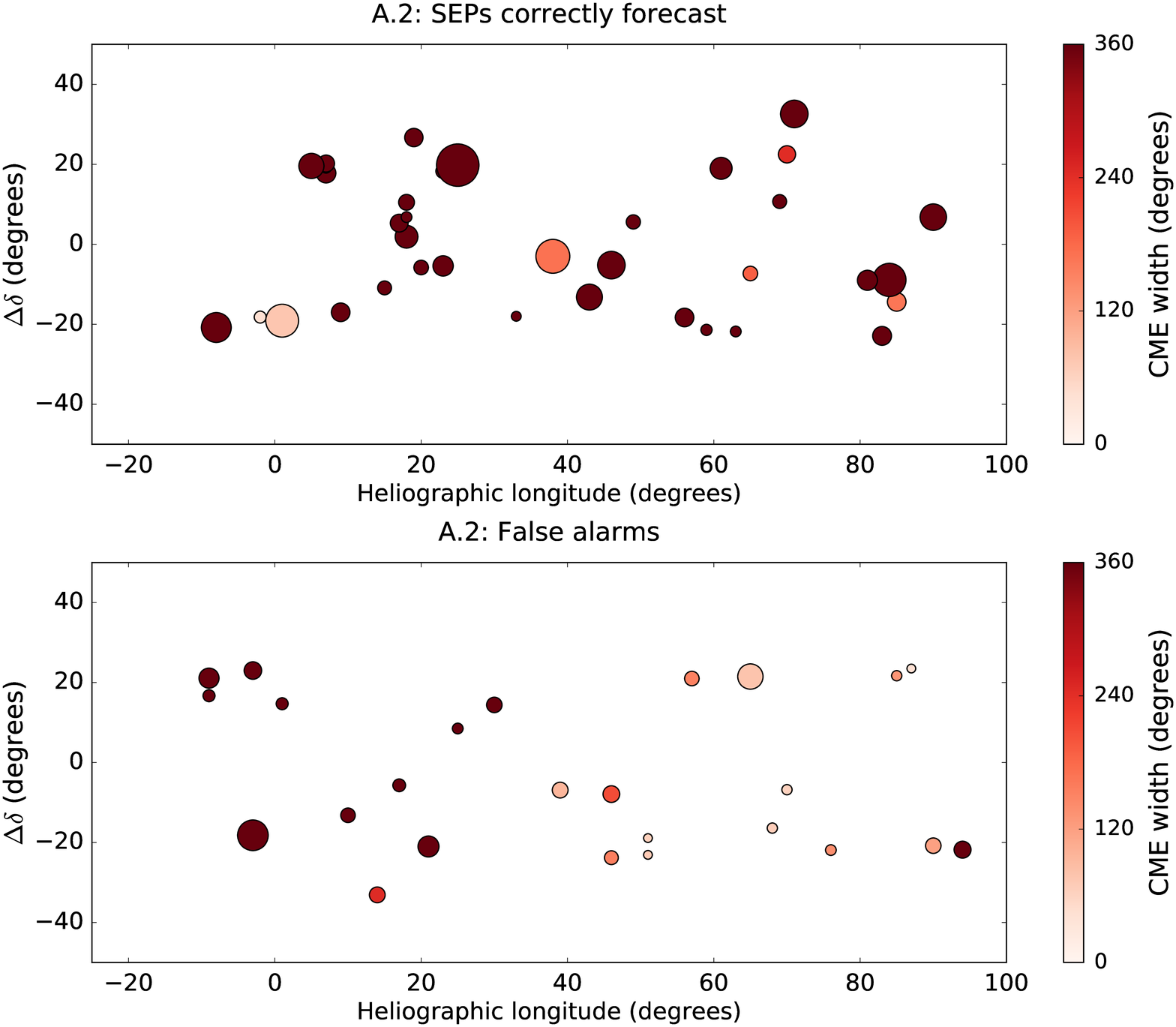}}
\caption{\(\Delta\delta\) \textit{versus} heliographic longitude for those western X-class flares which produced SEPs at Earth in time range 1 (\textit{top plot}); and for those which were false alarms according to algorithm A.2 (\textit{bottom plot}). The \textit{size of the marker} represents the relative duration of the flare: for example, the flare marked at S18W33 in the top plot had a duration of ten minutes, whereas the flare at S03W38 in the same plot lasted 120 minutes. The \textit{colour of the marker} represents CME width.}
\label{fig:alg_2_lon_v_delta_delta_width_duration}
\end{figure}

In the top plot of Figure~\ref{fig:alg_2_lon_v_delta_delta_width_duration} we plot \(\Delta\delta\) against the longitude of the 39 western X-class flares which produced an SEP event in time range 1. As in Figure~\ref{fig:alg1_lon_v_delta_delta_width_intensity}, the colour of the marker is representative of the width of the flare's associated CME as reported by CDAW, but in the case of Figure~\ref{fig:alg_2_lon_v_delta_delta_width_duration} the size of the marker reflects the duration of the flare itself. The bottom plot gives the same information, but for the false alarms according to algorithm A.2.

X-class flares which were false alarms tended to be shorter than those which produced SEPs. Average flare duration for the SEP-producing X-class flares was 46.3 minutes, and 25.6\% were longer than 60 minutes (``long duration flares''). For the false alarms, average flare duration was 24.9 minutes, and only 5.0\% (2/40) were long duration flares. It has previously been shown that there is an association between long duration flares and CMEs \citep{2006ApJ...650L.143Y}, therefore the trend with duration may be connected with the fact that large flares without CMEs are more likely to be false alarms.

In this case, too, the width of the associated CME is an important parameter. Of the 39 western X-class flares which produced SEPS at Earth, we were able definitively to associate 37 with a CME (the other two occurring during times when LASCO did not produce any data). Of those 37, 86.5\% (32/37) were halo CMEs. In contrast, for the false alarms, we were able to confirm associations with CMEs in 25 cases. Of these 25, only 44.0\% (11/25) were haloes.

\subsection{Algorithm A.2 over time range 2}

Over the longer period of time range 2, there were 197 western X-class flares which we analysed, and 39.1\% (77/197) produced SEPs at Earth. The false alarm rate was thus 60.9\% (120/197) and the algorithm failed to forecast 47.8\% (71/148) of SEP events. Of all the SEP events for which coordinates could be determined, it missed 55.0\% (94/171). Therefore the FAR was 0.61 and the CSI 0.29 without the missed eastern events, and 0.26 with them. The FAR is higher for this longer time period than that for time range 1. Appendix D provides the list of false alarms for the algorithm A.2 over time range 2. 

\begin{figure}
\centerline{\includegraphics[width=1\textwidth]{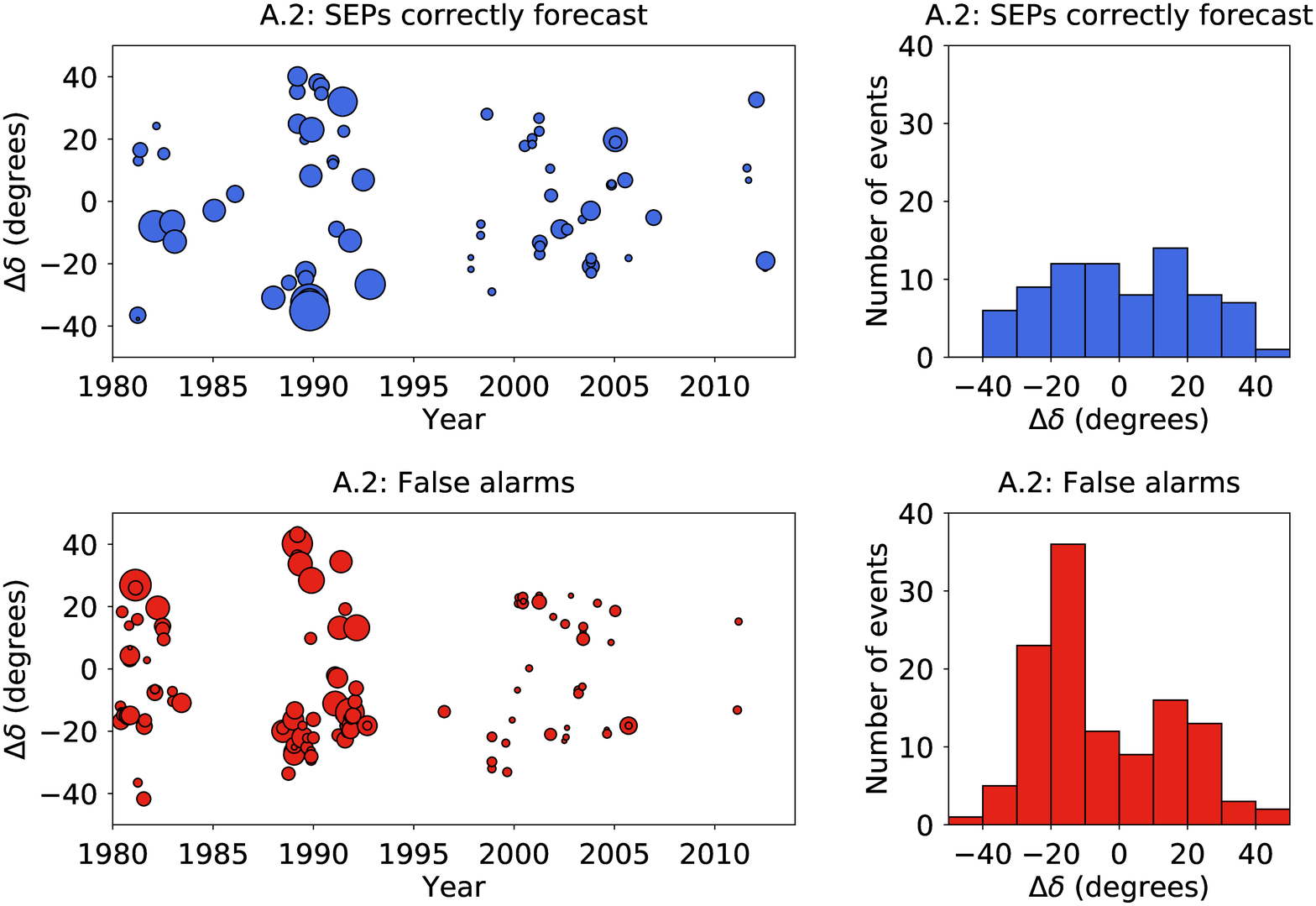}}
\caption{Plots of \(\Delta\delta\) against time for algorithm A.2 over time range 2, together with histograms of \(\Delta\delta\). The \textit{top plots} present the results for the solar events which were correctly forecast to produce SEPs at Earth (shown in \textit{blue}); the \textit{bottom plots} the false alarms (shown in \textit{red}). The \textit{size of the marker} in the \textit{left hand plots} is representative of the duration of the flare: for example, the flare in October 1989 shown at S35 in the top plot lasted 8 hours 48 minutes, whereas the flare in November 1998 shown at S29 in the same plot lasted 19 minutes}.
\label{fig:alg_2_date_v_dd_80}
\end{figure}

In Figure~\ref{fig:alg_2_date_v_dd_80} we plot \(\Delta\delta\) against date for this longer time period together with histograms for \(\Delta\delta\). In the left hand plots the duration of the flare is denoted by the size of the marker. Figure~\ref{fig:alg_2_date_v_dd_80} shows a significant difference in the \(\Delta\delta\) distribution for events which produced SEPs and false alarms. For the former the distribution is rather flat, whereas for the latter a high number of events are characterised by large \(\Delta\delta\).

There was a significantly higher number of false alarms from the southern solar hemisphere during Solar Cycle 22 (taken to be 1 January 1987 until 31 December 1995) (80\% - 40/50) than from the north (20\% - 10/50). Furthermore, in Solar Cycle 24 (taken to be from 1 January 2010 onwards) there were only two western X-class flares which were false alarms. 

It is also noted that X-class flares between 1980 and 1995 were, on average, longer than those \textit{post} 1995. It can be seen from Table~\ref{tab:instrs} that we have taken data from GOES 7 and its predecessors for dates before 1 March 1995, and from GOES 8 and its successors after that date. We are not aware of any reason why a change of instrument should produce such a result, nor are we aware of any change in the way flare duration has been measured.

\section{Improvement of the forecasting algorithms}

We examined ways in which the performance of the forecasting algorithms might be improved. We note in particular the following:

\begin{enumerate}
\item That algorithm A.1 produced the lowest number of false alarms, and that many of these had an associated flare intensity \(<\) M3.
\item That X-class flares without an associated CME, or associated with a CME of speed less than 500 km s\(^{-1}\), did not produce SEPs. 
\end{enumerate}

We therefore define a third forecasting algorithm as follows:

\begin{enumerate}[{A}.3]
\item A front-side CME with a reported speed of 1500 km s\(^{-1}\) or greater occurring west of E20 on the solar disk which is associated with a flare of class M3 or greater \textbf{or}
      
a solar flare of class X or greater which occurs west of E 20 on the solar disk and is associated with a CME of speed greater than 500 km s\(^{-1}\)

will result in an SEP event being detected at Earth.
\end{enumerate}

\begin{figure}
\centerline{\includegraphics[width=0.6\textwidth]{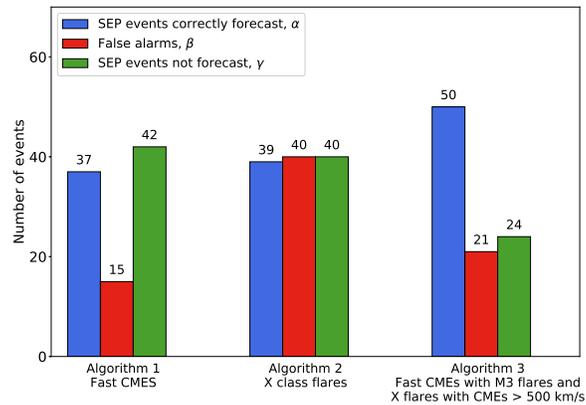}}
\caption{The numbers of correctly forecast SEP events, false alarms and SEP events which were not forecast for the three forecasting algorithms during time range 1.}
\label{fig:bar_plot_3}
\end{figure}

There were 71 such events in time range 1 and 70.4\% (50/71) produced SEPs at Earth. It should be noted that for this algorithm we have had to discard five of the SEP events which occurred during a time when there were no data from the LASCO coronagraph. Thus the false alarm rate was 29.6\% (21/71) and the algorithm missed 32.4\% (24/74) of SEP events for which the parent solar event was a western one, or 38.3\% (31/81) of all SEP events. The false alarm ratio is thus comparable to that produced by algorithm A.1, but A.3 misses far fewer SEP events and consequently the CSI is significantly higher at 0.53 not including the missed eastern events, or 0.49 were they to be included. The result is summarised in Table~\ref{tab:a3scores}. We also show the result graphically in Figure~\ref{fig:bar_plot_3} which is in the same format as Figure~\ref{fig:summary}. It may be possible to formulate better forecasting algorithms, but we suggest that increased forecasting accuracy will only come if the properties of both flares and CMEs are taken into account.

{\centering
\begin{table}
\caption{A summary of the evaluation scores for algorithm A.3 in the same format as Table~\ref{tab:scores}.}
\begin{tabular}{l c C{1cm} C{2.8cm} C{2.8cm}}
\toprule
Forecasting algorithm & Time range & FAR & CSI not including missed eastern events & CSI including missed eastern events \\
\midrule
A.3 & 1 & 0.30 & 0.53 & 0.49 \\
\bottomrule
\end{tabular}
\label{tab:a3scores}
\end{table}}

\section{Summary and conclusions}
\label{conclusion}

We have used historical data sets in order to assess the efficacy of two simple SEP forecasting algorithms which were based upon the occurrence of magnetically well-connected energetic solar events: western fast CMEs and X-class flares. We used in our definition of SEP event a threshold value for proton energy of \(>\)40 MeV.

An algorithm purely based on the detection of a fast CME (A.1) performs reasonably well in terms of false alarms (having a false alarm ratio of 28.8\%) but is missing a significant fraction of actual SEP events (53.1\%). It is unclear whether this is due to experimental limitations in the determination of the CME speed, or whether there are other physical properties which would need to be measured and included in the algorithm to assess the SEP producing potential of a CME more accurately. False alarms for this type of algorithm tend to be associated with flares of magnitude smaller than M3. There does not seem to be any positional trend in the source location of the false alarms.

An algorithm purely based on the detection of an intense flare (A.2) correctly forecasts almost the same number of SEP events as A.1 but has a much larger false alarm rate (50.6\%). Like A.1 it misses a significant fraction of SEP events (also 50.6\%). We found that false alarms for this algorithm tend to be flare events of shorter duration, compared to those which did produce SEPs. Of these false alarms, 37\% were not associated with a CME. An earlier study has analysed confined flares (CME-less flares) and emphasized that this kind of event tends not to produce SEPs \citep{2010SoPh..263..185K}. In terms of their longitudinal location, A.2 false alarm events were quite uniformly distributed. We also determined that SEP events not forecast by algorithm A.2 were preferentially located in the well-connected region (between W20 and W80), suggesting that for this region a lower flare magnitude threshold may need to be used.

When evaluated over a longer time range which includes Solar Cycle 21 (time range 2), algorithm A.2 performs less well than over time range 1. This may be due to instrumental effects associated with different GOES detectors being employed at different times, or it may be a real physical effect. We found that there is a systematic trend for flare durations to be larger in Cycle 22 compared with Cycle 23 and this may be an instrumental effect.

It has previously been suggested that the latitudinal separation, \(\Delta\delta\), between the flare location and the footpoint of the observing spacecraft plays a role in whether or not high-energy particles are detected \citep{2014EP&S...66..104G}. In our analysis, carried out over a larger time range, we found that false alarms for algorithm A.2 tended to be associated with a large latitudinal separation \(\Delta\delta\), whilst this was not the case for algorithm A.1. 

We defined a new forecasting algorithm, A.3, based upon the parameters of both flares and CMEs. This algorithm performed better than the algorithms based solely upon one type of solar event: it correctly forecast 70.4\% of SEP events during time range 1 and thus had a false alarm rate comparable to that of algorithm A.1 (29.6\%). It also missed many fewer SEP events (32.4\%, or 38.3\% if eastern events were to be included) than both algorithms A.1 and A.2. 

In test particle simulations it has been shown that SEPs may exhibit significant cross-field drift velocities depending on the configuration of the interplanetary magnetic field \citep{2013JGRA..118.5979D, 2013ApJ...774....4M}. Future work will assess whether the specific polarity of the magnetic field may influence whether or not SEPs were detected at a given location.

We have made available, in electronic form as supplementary material, lists of the \(>\)40 MeV proton false alarms according to each of the algorithms we analysed, together with a list of the solar events which produced the \(>\)40 MeV SEP events. We hope that these lists can be used as the basis for further studies and comparisons.

%

\acknowledgments

The CDAW catalogue is generated and maintained at the CDAW Data Center by NASA and The Catholic University of America in cooperation with the Naval Research Laboratory. The catalogue may be accessed at http://cdaw.gsfc.nasa.gov/CME\_list/index.html. The GOES SXR Flare List can be found at the Heliophysics Integrated Observatory website, http://hfe.helio-vo.eu, and we also used data as to heliographic coordinates from the SolarSoft Latest Events Flares List (gevloc) which can be obtained from the same site. Our proton intensity data were downloaded from the SEPEM website, http://dev.sepem.oma.be. We are very grateful to all those who maintain these data repositories as this work could not have been completed without them.

We have also used videos taken by the EIT instrument on board the SOHO spacecraft, and the AIA instrument on board the SDO spacecraft. SOHO is a project of international cooperation between ESA and NASA. We are grateful to them, and to the AIA, EVE, and HMI science teams.

Thanks, too, must go to the anonymous referee for their helpful comments. SD acknowledges support from the UK Science and Technology Facilities Council (STFC, grant ST/M00760X/1).

Disclosure of Potential Conflicts of Interest: the authors declare that they have no conflicts of interest.

\bibliographystyle{spr-mp-sola}

\begin{thebibliography}{41}
\ifx\bisbn     \undefined \def\bisbn  #1{ISBN #1}\fi
\ifx\binits    \undefined \def\binits#1{#1}\fi
\ifx\bauthor   \undefined \def\bauthor#1{#1}\fi
\ifx\batitle   \undefined \def\batitle#1{#1}\fi
\ifx\bjtitle   \undefined \def\bjtitle#1{\textit{#1}}\fi
\ifx\bvolume   \undefined \def\bvolume#1{\textbf{#1}}\fi
\ifx\byear     \undefined \def\byear#1{#1}\fi
\ifx\bissue    \undefined \def\bissue#1{#1}\fi
\ifx\bfpage    \undefined \def\bfpage#1{#1}\fi
\ifx\blpage    \undefined \def\blpage #1{#1}\fi
\ifx\burl      \undefined \def\burl#1{\textsf{#1}}\fi
\ifx\href      \undefined \def\href#1#2{\textsf{#2}}\fi
\ifx\betal     \undefined \def\betal{\textit{et al.}}\fi
\ifx\bctitle   \undefined \def\bctitle#1{#1}\fi
\ifx\beditor   \undefined \def\beditor#1{#1}\fi
\ifx\bbtitle   \undefined \def\bbtitle#1{\textit{#1}}\fi
\ifx\bedition  \undefined \def\bedition#1{#1}\fi
\ifx\bseriesno \undefined \def\bseriesno#1{\textbf{#1}}\fi
\ifx\blocation \undefined \def\blocation#1{#1}\fi
\ifx\bsertitle \undefined \def\bsertitle#1{\textit{#1}}\fi
\ifx\bsnm      \undefined \def\bsnm#1{#1}\fi
\ifx\bsuffix   \undefined \def\bsuffix#1{#1}\fi
\ifx\bparticle \undefined \def\bparticle#1{#1}\fi
\ifx\barticle  \undefined \def\barticle#1{}\fi
\ifx\binstitute  \undefined \def\binstitute#1{#1}\fi
\ifx\bpublisher  \undefined \def\bpublisher#1{#1}\fi
\ifx\doiurl    \undefined
  \def\doiurl#1{\href{http://dx.doi.org/#1}{\textsf{DOI}}}\fi
\ifx\arxivurl  \undefined
  \def\arxivurl#1{\href{http://arxiv.org/abs/#1}{\textsf{arXiv}}}\fi
\ifx\adsurl    \undefined
  \def\adsurl#1{\href{http://adsabs.harvard.edu/abs/#1}{\textsf{ADS}}}\fi
\ifx\botherref \undefined \def\botherref#1{}\fi
\ifx\url       \undefined \def\url#1{\textsf{#1}}\fi
\ifx\bchapter  \undefined \def\bchapter#1{}\fi
\ifx\bbook     \undefined \def\bbook#1{}\fi
\ifx\bcomment  \undefined \def\bcomment#1{#1}\fi
\ifx\oauthor   \undefined \def\oauthor#1{#1}\fi
\ifx\citeauthoryear \undefined\def \citeauthoryear#1{#1}\fi
\ifx\endbibitem\undefined \def\endbibitem{}\fi
\ifx\bconflocation  \undefined \def\bconflocation#1{#1} \fi

\bibitem[\protect\citeauthoryear{Alberti
  \textit{et~al.}}{2017}]{0004-637X-838-1-59}
\begin{barticle}
\bauthor{\bsnm{Alberti}, \binits{T.}},
\bauthor{\bsnm{Laurenza}, \binits{M.}},
\bauthor{\bsnm{Cliver}, \binits{E.W.}},
\bauthor{\bsnm{Storini}, \binits{M.}},
\bauthor{\bsnm{Consolini}, \binits{G.}},
\bauthor{\bsnm{Lepreti}, \binits{F.}}:
\byear{2017},
\batitle{Solar activity from 2006 to 2014 and short-term forecasts of solar
  proton events using the {E}{S}{P}{E}{R}{T}{A} model}.
\bjtitle{\apj}
\bvolume{838}(\bissue{1}),
\bfpage{59}.
\end{barticle}
\endbibitem

\bibitem[\protect\citeauthoryear{{Aran}, {Sanahuja}, and
  {Lario}}{2006}]{2006AdSpR..37.1240A}
\begin{barticle}
\bauthor{\bsnm{{Aran}}, \binits{A.}},
\bauthor{\bsnm{{Sanahuja}}, \binits{B.}},
\bauthor{\bsnm{{Lario}}, \binits{D.}}:
\byear{2006},
\batitle{{SOLPENCO: A solar particle engineering code}}.
\bjtitle{\asr}
\bvolume{37},
\bfpage{1240}.
\doiurl{10.1016/j.asr.2005.09.019}.
\adsurl{2006AdSpR..37.1240A}.
\end{barticle}
\endbibitem

\bibitem[\protect\citeauthoryear{Balch}{1999}]{1999RadiatMeas..30..231}
\begin{botherref}
\oauthor{\bsnm{Balch}, \binits{C.C.}}:
1999,
Sec proton prediction model: Verification and analysis.
\textit{\radmeas},
231.
\doiurl{10.1016/S1350-4487(99)00052-9}.
\end{botherref}
\endbibitem

\bibitem[\protect\citeauthoryear{{Balch}}{2008}]{2008SpWea...6.1001B}
\begin{barticle}
\bauthor{\bsnm{{Balch}}, \binits{C.C.}}:
\byear{2008},
\batitle{{Updated verification of the Space Weather Prediction Center's solar
  energetic particle prediction model}}.
\bjtitle{\sw}
\bvolume{6},
\bfpage{S01001}.
\doiurl{10.1029/2007SW000337}.
\adsurl{2008SpWea...6.1001B}.
\end{barticle}
\endbibitem

\bibitem[\protect\citeauthoryear{{Beck}
  \textit{et~al.}}{2005}]{2005AdSpR..36.1627B}
\begin{barticle}
\bauthor{\bsnm{{Beck}}, \binits{P.}},
\bauthor{\bsnm{{Latocha}}, \binits{M.}},
\bauthor{\bsnm{{Rollet}}, \binits{S.}},
\bauthor{\bsnm{{Stehno}}, \binits{G.}}:
\byear{2005},
\batitle{{TEPC reference measurements at aircraft altitudes during a solar
  storm}}.
\bjtitle{\asr}
\bvolume{36},
\bfpage{1627}.
\doiurl{10.1016/j.asr.2005.05.035}.
\adsurl{2005AdSpR..36.1627B}.
\end{barticle}
\endbibitem

\bibitem[\protect\citeauthoryear{{Belov}
  \textit{et~al.}}{2005}]{2005SoPh..229..135B}
\begin{barticle}
\bauthor{\bsnm{{Belov}}, \binits{A.}},
\bauthor{\bsnm{{Garcia}}, \binits{H.}},
\bauthor{\bsnm{{Kurt}}, \binits{V.}},
\bauthor{\bsnm{{Mavromichalaki}}, \binits{H.}},
\bauthor{\bsnm{{Gerontidou}}, \binits{M.}}:
\byear{2005},
\batitle{{Proton Enhancements and Their Relation to the X-Ray Flares During the
  Three Last Solar Cycles}}.
\bjtitle{\solphys}
\bvolume{229},
\bfpage{135}.
\doiurl{10.1007/s11207-005-4721-3}.
\adsurl{2005SoPh..229..135B}.
\end{barticle}
\endbibitem

\bibitem[\protect\citeauthoryear{{Bentley}
  \textit{et~al.}}{2011}]{2011AdSpR..47.2235B}
\begin{barticle}
\bauthor{\bsnm{{Bentley}}, \binits{R.D.}},
\bauthor{\bsnm{{Csillaghy}}, \binits{A.}},
\bauthor{\bsnm{{Aboudarham}}, \binits{J.}},
\bauthor{\bsnm{{Jacquey}}, \binits{C.}},
\bauthor{\bsnm{{Hapgood}}, \binits{M.A.}},
\bauthor{\bsnm{{Bocchialini}}, \binits{K.}},
\bauthor{\bsnm{{Messerotti}}, \binits{M.}},
\bauthor{\bsnm{{Brooke}}, \binits{J.}},
\bauthor{\bsnm{{Gallagher}}, \binits{P.}},
\bauthor{\bsnm{{Fox}}, \binits{P.}},
\bauthor{\bsnm{{Hurlburt}}, \binits{N.}},
\bauthor{\bsnm{{Roberts}}, \binits{D.A.}},
\bauthor{\bsnm{{Duarte}}, \binits{L.S.}}:
\byear{2011},
\batitle{{HELIO: The Heliophysics Integrated Observatory}}.
\bjtitle{\adv}
\bvolume{47},
\bfpage{2235}.
\doiurl{10.1016/j.asr.2010.02.006}.
\adsurl{2011AdSpR..47.2235B}.
\end{barticle}
\endbibitem

\bibitem[\protect\citeauthoryear{{Brueckner}
  \textit{et~al.}}{1995}]{1995SoPh..162..357B}
\begin{barticle}
\bauthor{\bsnm{{Brueckner}}, \binits{G.E.}},
\bauthor{\bsnm{{Howard}}, \binits{R.A.}},
\bauthor{\bsnm{{Koomen}}, \binits{M.J.}},
\bauthor{\bsnm{{Korendyke}}, \binits{C.M.}},
\bauthor{\bsnm{{Michels}}, \binits{D.J.}},
\bauthor{\bsnm{{Moses}}, \binits{J.D.}},
\bauthor{\bsnm{{Socker}}, \binits{D.G.}},
\bauthor{\bsnm{{Dere}}, \binits{K.P.}},
\bauthor{\bsnm{{Lamy}}, \binits{P.L.}},
\bauthor{\bsnm{{Llebaria}}, \binits{A.}},
\bauthor{\bsnm{{Bout}}, \binits{M.V.}},
\bauthor{\bsnm{{Schwenn}}, \binits{R.}},
\bauthor{\bsnm{{Simnett}}, \binits{G.M.}},
\bauthor{\bsnm{{Bedford}}, \binits{D.K.}},
\bauthor{\bsnm{{Eyles}}, \binits{C.J.}}:
\byear{1995},
\batitle{{The Large Angle Spectroscopic Coronagraph (LASCO)}}.
\bjtitle{\solphys}
\bvolume{162},
\bfpage{357}.
\doiurl{10.1007/BF00733434}.
\adsurl{1995SoPh..162..357B}.
\end{barticle}
\endbibitem

\bibitem[\protect\citeauthoryear{{Cliver}
  \textit{et~al.}}{2012}]{2012ApJ...756L..29C}
\begin{barticle}
\bauthor{\bsnm{{Cliver}}, \binits{E.W.}},
\bauthor{\bsnm{{Ling}}, \binits{A.G.}},
\bauthor{\bsnm{{Belov}}, \binits{A.}},
\bauthor{\bsnm{{Yashiro}}, \binits{S.}}:
\byear{2012},
\batitle{{Size distributions of solar flares and solar energetic particle
  events}}.
\bjtitle{\apjl}
\bvolume{756},
\bfpage{L29}.
\doiurl{10.1088/2041-8205/756/2/L29}.
\adsurl{2012ApJ...756L..29C}.
\end{barticle}
\endbibitem

\bibitem[\protect\citeauthoryear{{Crosby}
  \textit{et~al.}}{2010}]{2010cosp...38.4225C}
\begin{bchapter}
\bauthor{\bsnm{{Crosby}}, \binits{N.B.}},
\bauthor{\bsnm{{Glover}}, \binits{A.}},
\bauthor{\bsnm{{Aran}}, \binits{A.}},
\bauthor{\bsnm{{Bonnevie}}, \binits{C.}},
\bauthor{\bsnm{{Dyer}}, \binits{C.}},
\bauthor{\bsnm{{Gabriel}}, \binits{S.}},
\bauthor{\bsnm{{Hands}}, \binits{A.}},
\bauthor{\bsnm{{Heynderickx}}, \binits{D.}},
\bauthor{\bsnm{{Jacobs}}, \binits{C.}},
\bauthor{\bsnm{{Jiggens}}, \binits{P.}},
\bauthor{\bsnm{{King}}, \binits{D.}},
\bauthor{\bsnm{{Lawrence}}, \binits{G.}},
\bauthor{\bsnm{{Poedts}}, \binits{S.}},
\bauthor{\bsnm{{Sanahuja}}, \binits{B.}},
\bauthor{\bsnm{{Truscott}}, \binits{P.}}:
\byear{2010},
\bctitle{{SEPEM -Solar Energetic Particle Environment Modelling}}.
In: \bbtitle{38th COSPAR Scientific Assembly},
\bsertitle{COSPAR Meeting}
\bseriesno{38},
\bfpage{4}.
\adsurl{2010cosp...38.4225C}.
\end{bchapter}
\endbibitem

\bibitem[\protect\citeauthoryear{Dalla
  \textit{et~al.}}{2013}]{2013JGRA..118.5979D}
\begin{barticle}
\bauthor{\bsnm{Dalla}, \binits{S.}},
\bauthor{\bsnm{Marsh}, \binits{M.S.}},
\bauthor{\bsnm{Kelly}, \binits{J.}},
\bauthor{\bsnm{Laitinen}, \binits{T.}}:
\byear{2013},
\batitle{Solar energetic particle drifts in the parker spiral}.
\bjtitle{\jgr}
\bvolume{118}(\bissue{10}),
\bfpage{5979}.
\doiurl{10.1002/jgra.50589}.
\burl{http://dx.doi.org/10.1002/jgra.50589}.
\end{barticle}
\endbibitem

\bibitem[\protect\citeauthoryear{{Dierckxsens}
  \textit{et~al.}}{2015}]{2015SoPh..290..841D}
\begin{barticle}
\bauthor{\bsnm{{Dierckxsens}}, \binits{M.}},
\bauthor{\bsnm{{Tziotziou}}, \binits{K.}},
\bauthor{\bsnm{{Dalla}}, \binits{S.}},
\bauthor{\bsnm{{Patsou}}, \binits{I.}},
\bauthor{\bsnm{{Marsh}}, \binits{M.S.}},
\bauthor{\bsnm{{Crosby}}, \binits{N.B.}},
\bauthor{\bsnm{{Malandraki}}, \binits{O.}},
\bauthor{\bsnm{{Tsiropoula}}, \binits{G.}}:
\byear{2015},
\batitle{{Relationship between Solar Energetic Particles and Properties of
  Flares and CMEs: Statistical Analysis of Solar Cycle 23 Events}}.
\bjtitle{\solphys}
\bvolume{290},
\bfpage{841}.
\doiurl{10.1007/s11207-014-0641-4}.
\adsurl{2015SoPh..290..841D}.
\end{barticle}
\endbibitem

\bibitem[\protect\citeauthoryear{{Dumbovi{\'c}}
  \textit{et~al.}}{2015}]{2015SoPh..290..579D}
\begin{barticle}
\bauthor{\bsnm{{Dumbovi{\'c}}}, \binits{M.}},
\bauthor{\bsnm{{Devos}}, \binits{A.}},
\bauthor{\bsnm{{Vr{\v s}nak}}, \binits{B.}},
\bauthor{\bsnm{{Sudar}}, \binits{D.}},
\bauthor{\bsnm{{Rodriguez}}, \binits{L.}},
\bauthor{\bsnm{{Ru{\v z}djak}}, \binits{D.}},
\bauthor{\bsnm{{Leer}}, \binits{K.}},
\bauthor{\bsnm{{Vennerstr{\o}m}}, \binits{S.}},
\bauthor{\bsnm{{Veronig}}, \binits{A.}}:
\byear{2015},
\batitle{{Geoeffectiveness of Coronal Mass Ejections in the SOHO Era}}.
\bjtitle{\solphys}
\bvolume{290},
\bfpage{579}.
\doiurl{10.1007/s11207-014-0613-8}.
\adsurl{2015SoPh..290..579D}.
\end{barticle}
\endbibitem

\bibitem[\protect\citeauthoryear{{Feynman} and
  {Gabriel}}{2000}]{2000JGR...10510543F}
\begin{barticle}
\bauthor{\bsnm{{Feynman}}, \binits{J.}},
\bauthor{\bsnm{{Gabriel}}, \binits{S.B.}}:
\byear{2000},
\batitle{{On space weather consequences and predictions}}.
\bjtitle{\jgr}
\bvolume{105},
\bfpage{10543}.
\doiurl{10.1029/1999JA000141}.
\adsurl{2000JGR...10510543F}.
\end{barticle}
\endbibitem

\bibitem[\protect\citeauthoryear{{Garcia}}{1994}]{1994SoPh..154..275G}
\begin{barticle}
\bauthor{\bsnm{{Garcia}}, \binits{H.A.}}:
\byear{1994},
\batitle{{Temperature and emission measure from GOES soft X-ray measurements}}.
\bjtitle{\solphys}
\bvolume{154},
\bfpage{275}.
\doiurl{10.1007/BF00681100}.
\adsurl{1994SoPh..154..275G}.
\end{barticle}
\endbibitem

\bibitem[\protect\citeauthoryear{{Gopalswamy}
  \textit{et~al.}}{2009}]{2009EM&P..104..295G}
\begin{barticle}
\bauthor{\bsnm{{Gopalswamy}}, \binits{N.}},
\bauthor{\bsnm{{Yashiro}}, \binits{S.}},
\bauthor{\bsnm{{Michalek}}, \binits{G.}},
\bauthor{\bsnm{{Stenborg}}, \binits{G.}},
\bauthor{\bsnm{{Vourlidas}}, \binits{A.}},
\bauthor{\bsnm{{Freeland}}, \binits{S.}},
\bauthor{\bsnm{{Howard}}, \binits{R.}}:
\byear{2009},
\batitle{{The SOHO/LASCO CME Catalog}}.
\bjtitle{Earth, Moon, and Planets}
\bvolume{104},
\bfpage{295}.
\doiurl{10.1007/s11038-008-9282-7}.
\adsurl{2009EM\%26P..104..295G}.
\end{barticle}
\endbibitem

\bibitem[\protect\citeauthoryear{{Gopalswamy}
  \textit{et~al.}}{2014}]{2014EP&S...66..104G}
\begin{barticle}
\bauthor{\bsnm{{Gopalswamy}}, \binits{N.}},
\bauthor{\bsnm{{Xie}}, \binits{H.}},
\bauthor{\bsnm{{Akiyama}}, \binits{S.}},
\bauthor{\bsnm{{M{\"a}kel{\"a}}}, \binits{P.A.}},
\bauthor{\bsnm{{Yashiro}}, \binits{S.}}:
\byear{2014},
\batitle{{Major solar eruptions and high-energy particle events during solar
  cycle 24}}.
\bjtitle{Earth, Planets, and Space}
\bvolume{66},
\bfpage{104}.
\doiurl{10.1186/1880-5981-66-104}.
\adsurl{2014EP\%26S...66..104G}.
\end{barticle}
\endbibitem

\bibitem[\protect\citeauthoryear{{Grubb}}{1975}]{1975STIN...7628260G}
\begin{botherref}
\oauthor{\bsnm{{Grubb}}, \binits{R.N.}}:
1975,
{The SMS/GOES space environment monitor subsystem}.
\textit{NASA STI/Recon. Tech. Rep. N.}
\textbf{76}.
\adsurl{1975STIN...7628260G}.
\end{botherref}
\endbibitem

\bibitem[\protect\citeauthoryear{{Hargreaves}}{2005}]{2005AnGeo..23..359H}
\begin{barticle}
\bauthor{\bsnm{{Hargreaves}}, \binits{J.K.}}:
\byear{2005},
\batitle{{A new method of studying the relation between ionization rates and
  radio-wave absorption in polar-cap absorption events}}.
\bjtitle{\anng}
\bvolume{23},
\bfpage{359}.
\doiurl{10.5194/angeo-23-359-2005}.
\adsurl{2005AnGeo..23..359H}.
\end{barticle}
\endbibitem

\bibitem[\protect\citeauthoryear{{Harrison} and
  {Bewsher}}{2007}]{2007A&A...461.1155H}
\begin{barticle}
\bauthor{\bsnm{{Harrison}}, \binits{R.A.}},
\bauthor{\bsnm{{Bewsher}}, \binits{D.}}:
\byear{2007},
\batitle{{A benchmark event sequence for mass ejection onset studies. A flare
  associated CME with coronal dimming, ascending pre-flare loops and a
  transient cool loop}}.
\bjtitle{\aap}
\bvolume{461},
\bfpage{1155}.
\doiurl{10.1051/0004-6361:20066100}.
\adsurl{2007A\%26A...461.1155H}.
\end{barticle}
\endbibitem

\bibitem[\protect\citeauthoryear{{Hoff}, {Townsend}, and
  {Zapp}}{2004}]{2004AdSpR..34.1347H}
\begin{barticle}
\bauthor{\bsnm{{Hoff}}, \binits{J.L.}},
\bauthor{\bsnm{{Townsend}}, \binits{L.W.}},
\bauthor{\bsnm{{Zapp}}, \binits{E.N.}}:
\byear{2004},
\batitle{{Interplanetary crew doses and dose equivalents: variations among
  different bone marrow and skin sites}}.
\bjtitle{\asr}
\bvolume{34},
\bfpage{1347}.
\doiurl{10.1016/j.asr.2003.08.056}.
\adsurl{2004AdSpR..34.1347H}.
\end{barticle}
\endbibitem

\bibitem[\protect\citeauthoryear{Kahler and Ling}{2015}]{SWE:SWE20256}
\begin{barticle}
\bauthor{\bsnm{Kahler}, \binits{S.W.}},
\bauthor{\bsnm{Ling}, \binits{A.}}:
\byear{2015},
\batitle{Dynamic sep event probability forecasts}.
\bjtitle{\sw}
\bvolume{13}(\bissue{10}),
\bfpage{665}.
\bcomment{2015SW001222}.
\doiurl{10.1002/2015SW001222}.
\end{barticle}
\endbibitem

\bibitem[\protect\citeauthoryear{{Kahler}, {Cliver}, and
  {Ling}}{2007}]{2007JASTP..69...43K}
\begin{barticle}
\bauthor{\bsnm{{Kahler}}, \binits{S.W.}},
\bauthor{\bsnm{{Cliver}}, \binits{E.W.}},
\bauthor{\bsnm{{Ling}}, \binits{A.G.}}:
\byear{2007},
\batitle{{Validating the proton prediction system (PPS)}}.
\bjtitle{\jastp}
\bvolume{69},
\bfpage{43}.
\doiurl{10.1016/j.jastp.2006.06.009}.
\adsurl{2007JASTP..69...43K}.
\end{barticle}
\endbibitem

\bibitem[\protect\citeauthoryear{{Klein}, {Trottet}, and
  {Klassen}}{2010}]{2010SoPh..263..185K}
\begin{barticle}
\bauthor{\bsnm{{Klein}}, \binits{K.-L.}},
\bauthor{\bsnm{{Trottet}}, \binits{G.}},
\bauthor{\bsnm{{Klassen}}, \binits{A.}}:
\byear{2010},
\batitle{{Energetic Particle Acceleration and Propagation in Strong CME-Less
  Flares}}.
\bjtitle{\solphys}
\bvolume{263},
\bfpage{185}.
\doiurl{10.1007/s11207-010-9540-5}.
\adsurl{2010SoPh..263..185K}.
\end{barticle}
\endbibitem

\bibitem[\protect\citeauthoryear{{Klein}
  \textit{et~al.}}{2011}]{2011SoPh..269..309K}
\begin{barticle}
\bauthor{\bsnm{{Klein}}, \binits{K.-L.}},
\bauthor{\bsnm{{Trottet}}, \binits{G.}},
\bauthor{\bsnm{{Samwel}}, \binits{S.}},
\bauthor{\bsnm{{Malandraki}}, \binits{O.}}:
\byear{2011},
\batitle{{Particle Acceleration and Propagation in Strong Flares without Major
  Solar Energetic Particle Events}}.
\bjtitle{\solphys}
\bvolume{269},
\bfpage{309}.
\doiurl{10.1007/s11207-011-9710-0}.
\adsurl{2011SoPh..269..309K}.
\end{barticle}
\endbibitem

\bibitem[\protect\citeauthoryear{{Kwon}, {Zhang}, and
  {Vourlidas}}{2015}]{2015ApJ...799L..29K}
\begin{barticle}
\bauthor{\bsnm{{Kwon}}, \binits{R.-Y.}},
\bauthor{\bsnm{{Zhang}}, \binits{J.}},
\bauthor{\bsnm{{Vourlidas}}, \binits{A.}}:
\byear{2015},
\batitle{{Are Halo-like Solar Coronal Mass Ejections Merely a Matter of
  Geometric Projection Effects?}}
\bjtitle{\apjl}
\bvolume{799},
\bfpage{L29}.
\doiurl{10.1088/2041-8205/799/2/L29}.
\adsurl{2015ApJ...799L..29K}.
\end{barticle}
\endbibitem

\bibitem[\protect\citeauthoryear{{Laurenza}
  \textit{et~al.}}{2009}]{2009SpWea...7.4008L}
\begin{barticle}
\bauthor{\bsnm{{Laurenza}}, \binits{M.}},
\bauthor{\bsnm{{Cliver}}, \binits{E.W.}},
\bauthor{\bsnm{{Hewitt}}, \binits{J.}},
\bauthor{\bsnm{{Storini}}, \binits{M.}},
\bauthor{\bsnm{{Ling}}, \binits{A.G.}},
\bauthor{\bsnm{{Balch}}, \binits{C.C.}},
\bauthor{\bsnm{{Kaiser}}, \binits{M.L.}}:
\byear{2009},
\batitle{{A technique for short-term warning of solar energetic particle events
  based on flare location, flare size, and evidence of particle escape}}.
\bjtitle{\sw}
\bvolume{7},
\bfpage{4008}.
\doiurl{10.1029/2007SW000379}.
\adsurl{2009SpWea...7.4008L}.
\end{barticle}
\endbibitem

\bibitem[\protect\citeauthoryear{{Luhmann}
  \textit{et~al.}}{2010}]{2010AdSpR..46....1L}
\begin{barticle}
\bauthor{\bsnm{{Luhmann}}, \binits{J.G.}},
\bauthor{\bsnm{{Ledvina}}, \binits{S.A.}},
\bauthor{\bsnm{{Odstrcil}}, \binits{D.}},
\bauthor{\bsnm{{Owens}}, \binits{M.J.}},
\bauthor{\bsnm{{Zhao}}, \binits{X.-P.}},
\bauthor{\bsnm{{Liu}}, \binits{Y.}},
\bauthor{\bsnm{{Riley}}, \binits{P.}}:
\byear{2010},
\batitle{{Cone model-based SEP event calculations for applications to
  multipoint observations}}.
\bjtitle{\asr}
\bvolume{46},
\bfpage{1}.
\doiurl{10.1016/j.asr.2010.03.011}.
\adsurl{2010AdSpR..46....1L}.
\end{barticle}
\endbibitem

\bibitem[\protect\citeauthoryear{{Marqu{\'e}}, {Posner}, and
  {Klein}}{2006}]{2006ApJ...642.1222M}
\begin{barticle}
\bauthor{\bsnm{{Marqu{\'e}}}, \binits{C.}},
\bauthor{\bsnm{{Posner}}, \binits{A.}},
\bauthor{\bsnm{{Klein}}, \binits{K.-L.}}:
\byear{2006},
\batitle{{Solar Energetic Particles and Radio-silent Fast Coronal Mass
  Ejections}}.
\bjtitle{\apj}
\bvolume{642},
\bfpage{1222}.
\doiurl{10.1086/501157}.
\adsurl{2006ApJ...642.1222M}.
\end{barticle}
\endbibitem

\bibitem[\protect\citeauthoryear{{Marsh}
  \textit{et~al.}}{2013}]{2013ApJ...774....4M}
\begin{barticle}
\bauthor{\bsnm{{Marsh}}, \binits{M.S.}},
\bauthor{\bsnm{{Dalla}}, \binits{S.}},
\bauthor{\bsnm{{Kelly}}, \binits{J.}},
\bauthor{\bsnm{{Laitinen}}, \binits{T.}}:
\byear{2013},
\batitle{{Drift-induced Perpendicular Transport of Solar Energetic Particles}}.
\bjtitle{\apj}
\bvolume{774},
\bfpage{4}.
\doiurl{10.1088/0004-637X/774/1/4}.
\adsurl{2013ApJ...774....4M}.
\end{barticle}
\endbibitem

\bibitem[\protect\citeauthoryear{Marsh \textit{et~al.}}{2015}]{SWE:SWE20233}
\begin{barticle}
\bauthor{\bsnm{Marsh}, \binits{M.S.}},
\bauthor{\bsnm{Dalla}, \binits{S.}},
\bauthor{\bsnm{Dierckxsens}, \binits{M.}},
\bauthor{\bsnm{Laitinen}, \binits{T.}},
\bauthor{\bsnm{Crosby}, \binits{N.B.}}:
\byear{2015},
\batitle{{S}{P}{A}{R}{X}: A modeling system for solar energetic particle
  radiation space weather forecasting}.
\bjtitle{\sw}
\bvolume{13}(\bissue{6}),
\bfpage{386}.
\bcomment{2014SW001120}.
\doiurl{10.1002/2014SW001120}.
\end{barticle}
\endbibitem

\bibitem[\protect\citeauthoryear{Papaioannou
  \textit{et~al.}}{2015}]{1742-6596-632-1-012075}
\begin{barticle}
\bauthor{\bsnm{Papaioannou}, \binits{A.}},
\bauthor{\bsnm{Anastasiadis}, \binits{A.}},
\bauthor{\bsnm{Sandberg}, \binits{I.}},
\bauthor{\bsnm{Georgoulis}, \binits{M.K.}},
\bauthor{\bsnm{Tsiropoula}, \binits{G.}},
\bauthor{\bsnm{Tziotziou}, \binits{K.}},
\bauthor{\bsnm{Jiggens}, \binits{P.}},
\bauthor{\bsnm{Hilgers}, \binits{A.}}:
\byear{2015},
\batitle{A novel forecasting system for solar particle events and flares
  ({F}{O}{R}{S}{P}{E}{F})}.
\bjtitle{\jpcs}
\bvolume{632}(\bissue{1}),
\bfpage{012075}.
\end{barticle}
\endbibitem

\bibitem[\protect\citeauthoryear{{Papaioannou}
  \textit{et~al.}}{2016}]{2016JSWSC...6A..42P}
\begin{barticle}
\bauthor{\bsnm{{Papaioannou}}, \binits{A.}},
\bauthor{\bsnm{{Sandberg}}, \binits{I.}},
\bauthor{\bsnm{{Anastasiadis}}, \binits{A.}},
\bauthor{\bsnm{{Kouloumvakos}}, \binits{A.}},
\bauthor{\bsnm{{Georgoulis}}, \binits{M.K.}},
\bauthor{\bsnm{{Tziotziou}}, \binits{K.}},
\bauthor{\bsnm{{Tsiropoula}}, \binits{G.}},
\bauthor{\bsnm{{Jiggens}}, \binits{P.}},
\bauthor{\bsnm{{Hilgers}}, \binits{A.}}:
\byear{2016},
\batitle{{Solar flares, coronal mass ejections and solar energetic particle
  event characteristics}}.
\bjtitle{\swsc}
\bvolume{6}(\bissue{27}),
\bfpage{A42}.
\doiurl{10.1051/swsc/2016035}.
\adsurl{2016JSWSC...6A..42P}.
\end{barticle}
\endbibitem

\bibitem[\protect\citeauthoryear{{Park}, {Moon}, and
  {Gopalswamy}}{2012}]{2012JGRA..117.8108P}
\begin{barticle}
\bauthor{\bsnm{{Park}}, \binits{J.}},
\bauthor{\bsnm{{Moon}}, \binits{Y.-J.}},
\bauthor{\bsnm{{Gopalswamy}}, \binits{N.}}:
\byear{2012},
\batitle{{Dependence of solar proton events on their associated activities:
  Coronal mass ejection parameters}}.
\bjtitle{\jgr}
\bvolume{117},
\bfpage{A08108}.
\doiurl{10.1029/2011JA017477}.
\adsurl{2012JGRA..117.8108P}.
\end{barticle}
\endbibitem

\bibitem[\protect\citeauthoryear{Posner}{2007}]{SWE:SWE185}
\begin{barticle}
\bauthor{\bsnm{Posner}, \binits{A.}}:
\byear{2007},
\batitle{Up to 1-hour forecasting of radiation hazards from solar energetic ion
  events with relativistic electrons}.
\bjtitle{\sw}
\bvolume{5}(\bissue{5}),
\bfpage{n/a}.
\bcomment{S05001}.
\doiurl{10.1029/2006SW000268}.
\end{barticle}
\endbibitem

\bibitem[\protect\citeauthoryear{{Reinard} and
  {Andrews}}{2006}]{2006AdSpR..38..480R}
\begin{barticle}
\bauthor{\bsnm{{Reinard}}, \binits{A.A.}},
\bauthor{\bsnm{{Andrews}}, \binits{M.A.}}:
\byear{2006},
\batitle{{Comparison of CME characteristics for SEP and non-SEP related
  events}}.
\bjtitle{\asr}
\bvolume{38},
\bfpage{480}.
\doiurl{10.1016/j.asr.2005.01.028}.
\adsurl{2006AdSpR..38..480R}.
\end{barticle}
\endbibitem

\bibitem[\protect\citeauthoryear{{Smart} and
  {Shea}}{1989}]{1989AdSpR...9..281S}
\begin{barticle}
\bauthor{\bsnm{{Smart}}, \binits{D.F.}},
\bauthor{\bsnm{{Shea}}, \binits{M.A.}}:
\byear{1989},
\batitle{{PPS-87 - A new event oriented solar proton prediction model}}.
\bjtitle{\asr}
\bvolume{9},
\bfpage{281}.
\doiurl{10.1016/0273-1177(89)90450-X}.
\adsurl{1989AdSpR...9..281S}.
\end{barticle}
\endbibitem

\bibitem[\protect\citeauthoryear{{Vr{\v s}nak}, {Sudar}, and {Ru{\v
  z}djak}}{2005}]{2005A&A...435.1149V}
\begin{barticle}
\bauthor{\bsnm{{Vr{\v s}nak}}, \binits{B.}},
\bauthor{\bsnm{{Sudar}}, \binits{D.}},
\bauthor{\bsnm{{Ru{\v z}djak}}, \binits{D.}}:
\byear{2005},
\batitle{{The CME-flare relationship: Are there really two types of CMEs?}}
\bjtitle{\aap}
\bvolume{435},
\bfpage{1149}.
\doiurl{10.1051/0004-6361:20042166}.
\adsurl{2005A\%26A...435.1149V}.
\end{barticle}
\endbibitem

\bibitem[\protect\citeauthoryear{{Wang} and
  {Zhang}}{2007}]{2007ApJ...665.1428W}
\begin{barticle}
\bauthor{\bsnm{{Wang}}, \binits{Y.}},
\bauthor{\bsnm{{Zhang}}, \binits{J.}}:
\byear{2007},
\batitle{{A Comparative Study between Eruptive X-Class Flares Associated with
  Coronal Mass Ejections and Confined X-Class Flares}}.
\bjtitle{\apj}
\bvolume{665},
\bfpage{1428}.
\doiurl{10.1086/519765}.
\adsurl{2007ApJ...665.1428W}.
\end{barticle}
\endbibitem

\bibitem[\protect\citeauthoryear{Winter and
  Ledbetter}{2015}]{0004-637X-809-1-105}
\begin{barticle}
\bauthor{\bsnm{Winter}, \binits{L.M.}},
\bauthor{\bsnm{Ledbetter}, \binits{K.}}:
\byear{2015},
\batitle{Type ii and type iii radio bursts and their correlation with solar
  energetic proton events}.
\bjtitle{\apj}
\bvolume{809}(\bissue{1}),
\bfpage{105}.
\burl{http://stacks.iop.org/0004-637X/809/i=1/a=105}.
\end{barticle}
\endbibitem

\bibitem[\protect\citeauthoryear{{Yashiro}
  \textit{et~al.}}{2006}]{2006ApJ...650L.143Y}
\begin{barticle}
\bauthor{\bsnm{{Yashiro}}, \binits{S.}},
\bauthor{\bsnm{{Akiyama}}, \binits{S.}},
\bauthor{\bsnm{{Gopalswamy}}, \binits{N.}},
\bauthor{\bsnm{{Howard}}, \binits{R.A.}}:
\byear{2006},
\batitle{{Different Power-Law Indices in the Frequency Distributions of Flares
  with and without Coronal Mass Ejections}}.
\bjtitle{\apjl}
\bvolume{650},
\bfpage{L143}.
\doiurl{10.1086/508876}.
\adsurl{2006ApJ...650L.143Y}.
\end{barticle}
\endbibitem

\end{thebibliography}

\newpage
\appendix
\section{Association of Solar Flares and CMEs}
\label{app:method}

It has long been accepted that solar flares and CMEs, particularly energetic events, often occur within a short time of each other from the same solar active region, but making associations between them is no trivial exercise. There is no standard approach: for example, \citealp{2006AdSpR..38..480R} associate a flare with a CME if the CME occurred within a 2 hour window centred on the time of the peak of the flare; others make associations by using both temporal and spatial criteria \citep{2005A&A...435.1149V, 2015SoPh..290..579D}. Below we describe a method of making associations between CMEs and flares automatically, and evaluate its accuracy. 

In the light of the connection between high energy eruptive events and SEPs we decided to look for associations involving CMEs reported by CDAW to have a speed of 1000 km s\(^{-1}\) or faster (``rapid CMEs"), and flares reported in the GOES SXR list to be of class M5 or greater (``intense flares"). We examined all such events between 1 July 2011 and 31 August 2012, that period being chosen solely because it provided a data set which was small enough to allow individual observation of each event, yet large enough to allow wider conclusions to be drawn. 

There were 55 rapid CMEs and 32 intense flares reported in the 13 month period under investigation. Of these, we did not study further 3 of the rapid CMEs and 1 of the intense flares because they coincided with data gaps. Hence there were 83 events which formed the basis of our study of flare\textendash CME associations. 

In order to set a benchmark against which any automated method of associating CMEs and flares could be judged, we needed to know definitively whether any of the 83 energetic events were associated with another solar event. Consequently we watched movies at 193 \r{A} of each one of these events, each movie having been created from data obtained by the \textit{Atmospheric Imaging Assembly} (AIA) on board the \textit{Solar Dynamics Observatory} (SDO) spacecraft. 

For each of the intense flares, identification was done visually from the AIA / SDO movies. We looked for increases in intensity on the solar surface at the time of the flare specified by the GOES SXR list, but in cases where the site of the flare was not obvious we accepted the reported coordinates. Whilst watching the movies of the intense flares, we also searched for evidence of an associated CME. If we were able to see any ejected material, any loop distortion or any coronal dimming consistent with the flare site within 1 hour either side of the reported time of the flare, we associated that flare with a CME (whether or not this was a rapid CME).

Rapid CMEs were identified by searching visually for evidence of any ejected material, any loop distortion or any coronal dimming at the time reported by CDAW. If such evidence was present (and was consistent with a front-side event), the CME was regarded as having occurred on the face of the disk; if there was no such evidence, the CME was regarded as a back-side event. Associations were made between a rapid CME and a flare (whether or not this was an intense flare) if the reported time of the CME (\textit{i.e.} the time the CME was first seen in the LASCO C2 images) fell between 1 hour before the reported start of the flare and 1 hour after its reported end, and the evidence of the CME was consistent with the flare site.

As a result of making the associations manually we found that 35 of the 52 fast CMEs were on the face of the disk. This proportion is slightly higher than might have been expected (given that we can only see one side of the Sun at any one time, we might expect that only half of the CMEs we see would be from the face of the disk), but can be explained by two factors: first, there were large numbers of CMEs from same active regions (two active regions produced five each, and one other eight) and this may slightly distort the figures; secondly 17 of the 52 events were reported to occur very close to the limb, meaning that we may have seen a CME which originated from just behind the limb.

Of the 35 rapid CMEs which occurred on the face of the disk, all were associated with a flare of some kind; 46\% (16/35) were associated with an intense flare.  Of the 31 intense flares, 84\% (26/31) were associated with a CME. 

In every instance where we had associated a solar flare with a CME, the flare was reported in the GOES SXR list as having commenced before the CME was first reported in the CDAW catalogue. It should be noted that this is not an indication of actual chronology - as an example of where there is evidence of a CME lifting off before its associated flare, see \citealp{2007A&A...461.1155H} - but it is of significance when devising a method of automatically making associations between flares and CMEs.

CDAW reports the time of a CME as being when it is first seen in images produced by the LASCO C2 coronagraph. This instrument, however, has a field of view between about 2 and 6 solar radii (as measured from the Sun's centre) and the images used by CDAW have a cadence of, at best, 12 minutes and sometimes much longer. The combination of these factors means that the reported time of the CME may be many minutes after is actual ``lift-off" time, \(t_{o}\). 

Any attempt to make an estimate of \(t_{o}\) faces a number of difficulties: there is no information as to the height of the CME when it was first ejected; no information as to whether it has accelerated or decelerated before its first appearance in the C2 images; and no information as to the direction of the CME. Nevertheless, finding a first approximation of \(t_{o}\) is more likely to result in accurate associations between CMEs and flares than using the time of the CME as reported by CDAW.

We make the simple assumptions that by the time the CME reaches the field of view of the C2 coronagraph it has travelled (at least) one solar radius and has undergone neither significant acceleration nor deceleration. An estimate for \(t_{o}\) is then obtained by using the reported speed of the CME.

In order to take into account of the difficulties caused by the cadence of the images, we define \(\Delta\)t as a number of minutes both before and after a flare. For example, if we take \(\Delta\)t = 12, we compare \(t_{o}\) with a time window opening 12 minutes before the flare began and closing 12 minutes after it ended. Plainly, the greater \(\Delta\)t, the more likely it is that \(t_{o}\) will fall within the window, and hence the greater the likelihood of false associations being made.

We found that a good correlation could be found between those flare\textendash  CME associations which had made manually and those using a value of \(\Delta\)t of just 30 minutes. We did investigate whether it may be possible to improve the accuracy of the method by imposing a spatial criterion, for example by requiring the position angle of the CME to agree with the latitude and longitude of the flare to within a particular number of degrees. In fact we found that overall accuracy was not improved by the imposition of such a criterion.

There will, of course, always be a small number of (usually) false associations when using this automatic method given that occasionally apparently unconnected solar events sometimes occur almost simultaneously. Nevertheless, in our sample the method correctly identified 98\% (60/61) associations and correctly identified 86\% (19/22) non-associations, an overall success rate in 95\% (79/83) of cases.

\newpage
\begin{landscape}
\begin{center}
\section{Algorithm 1: False alarms in time range 1}
\label{app:alg1falist}
\begin{table}[h!]
\caption{List of fast CMEs between 1 January 1996 and 31 March 2013 which were false alarms. Column 1 gives the time the CME was first reported, column 2 its speed, and column 3 its acceleration all as reported by CDAW. Column 4 is the position angle at which the height-time measurements had been made (called by CDAW the ``measurement position angle"). Column 5 is the CME width. Columns 6 to 10 give the parameters of the associated flare: its start time, heliographic latitude, heliographic longitude, class, and duration.}
\begin{tabular}
{l 
>{\centering\arraybackslash}m{1.2cm} 
>{\centering\arraybackslash}m{1.2cm} 
>{\centering\arraybackslash}m{1.2cm} 
>{\centering\arraybackslash}m{1.2cm} 
c c c c 
>{\centering\arraybackslash}m{2cm} }

\addlinespace
\toprule

   \multicolumn{5}{c}{CME parameters} & 
   \multicolumn{5}{c}{Flare parameters} \\
  First reported  &
  \makecell{V \\ (km s\(^{-1}\))} &
  \makecell{Accel. \\ (km s\(^{-2}\))} &
  Mpa (degrees) &
  Width (degrees) &
  Time start &
  Lat &
  Lon &
  Class &
  \makecell{Duration \\ (hrs: mins)}  \\
\midrule
  1998-11-24T02:30 & 1798 & -12.5 & 225 & 360 & 1998-11-24T02:07 & -20 & 94 & X1.0 & 00:30\\
  2000-01-06T07:31 & 1813 & 10.7 & 342 & 67 & 2000-01-06T06:45 & 24 & 35 & C5.8 & 00:21\\
  2000-05-05T15:50 & 1594 & -103.4 & 265 & 360 & 2000-05-05T15:19 & -15 & 97 & M1.5 & 02:09\\
  2000-06-25T07:54 & 1617 & -17.5 & 274 & 165 & 2000-06-25T07:17 & 16 & 55 & M1.9 & 01:04\\
  2001-07-19T10:30 & 1668 & -11.6 & 252 & 166 & 2001-07-19T09:52 & -8 & 62 & M1.8 & 00:25\\
  2002-03-22T11:06 & 1750 & -22.5 & 259 & 360 & 2002-03-22T10:12 & -10 & 95 & M1.6 & 01:40\\
  2002-05-30T05:06 & 1625 & 67.0 & 275 & 144 & 2002-05-30T04:24 & 5 & 95 & M1.3 & 01:49\\
  2002-08-16T12:30 & 1585 & -67.1 & 121 & 360 & 2002-08-16T11:32 & -14 & -20 & M5.2 & 01:35\\
  2003-03-18T12:30 & 1601 & -13.3 & 266 & 209 & 2003-03-18T11:51 & -15 & 46 & X1.5 & 00:29\\
  2003-06-02T00:30 & 1656 & 42.5 & 248 & 172 & 2003-06-02T00:07 & -8 & 89 & M6.5 & 00:36\\
  2003-11-18T08:50 & 1660 & -3.3 & 206 & 360 & 2003-11-18T08:12 & -2 & -18 & M3.9 & 00:47\\
  2004-04-11T04:30 & 1645 & -77.6 & 237 & 314 & 2004-04-11T03:54 & -16 & 46 & C9.6 & 00:41\\
  2005-07-09T22:30 & 1540 & -168.5 & 328 & 360 & 2005-07-09T21:47 & 12 & 28 & M2.8 & 00:32\\
  2005-09-13T20:00 & 1866 & 11.5 & 149 & 360 & 2005-09-13T19:19 & -11 & -3 & X1.5 & 01:38\\
  2013-02-06T00:24 & 1867 & -8.2 & 31 & 271 & 2013-02-06T00:04 & 22 & -19 & C8.7 & 00:37\\
\bottomrule
\end{tabular}
\end{table}
\end{center}
\end{landscape}

\newpage

\begin{landscape}
\begin{center}
\section{Algorithm 2: False alarms in time range 1}
\label{app:alg2falist}
\begin{table}[h!]
\caption{List of X-class flares between 1 January 1996 and 31 March 2013 which were false alarms. Column 1 gives the time the CME was first reported, column 2 its speed, and column 3 its acceleration all as reported by CDAW. Column 4 is the position angle at which the height-time measurements had been made (called by CDAW the ``measurement position angle''). Column 5 is the CME width. Columns 6 to 10 give the parameters of the associated flare: its start time, heliographic latitude, heliographic longitude, class, and duration.} 
\begin{tabular}
{l 
>{\centering\arraybackslash}m{1.2cm} 
>{\centering\arraybackslash}m{1.2cm} 
>{\centering\arraybackslash}m{1.2cm} 
>{\centering\arraybackslash}m{1.2cm} 
c c c c 
>{\centering\arraybackslash}m{2cm} }

\addlinespace
\toprule

   \multicolumn{5}{c}{CME parameters} & 
   \multicolumn{5}{c}{Flare parameters} \\
  First reported &
  \makecell{V \\ (km s\(^{-1}\))} &
  \makecell{Accel. \\ (km s\(^{-2}\))} &
  Mpa (degrees) &
  Width (degrees) &
  Time start &
  Lat &
  Lon &
  Class &
  \makecell{Duration \\ (hrs: mins)}  \\
\midrule

  LASCO data gap &  &  &  &  & 1996-07-09T09:01 & -10 & 30 & X2.6 & 00:48\\
  LASCO data gap &  &  &  &  & 1998-11-22T16:10 & -30 & 89 & X2.5 & 00:22\\
  LASCO data gap &  &  &  &  & 1998-11-23T06:28 & -28 & 89 & X2.2 & 00:30\\
  1998-11-24T02:30 & 1798 & -12.5 & 225 & 360 & 1998-11-24T02:07 & -20 & 94 & X1.0 & 00:30\\
  1999-08-02T22:26 & 292 & 0.9 & 264 & 157 & 1999-08-02T21:18 & -18 & 46 & X1.4 & 00:20\\
  1999-08-28T18:26 & 462 & 1.1 & 221 & 245 & 1999-08-28T17:52 & -26 & 14 & X1.1 & 00:26\\
  1999-11-27T12:54 & 235 & 64.2 & 256 & 68 & 1999-11-27T12:05 & -15 & 68 & X1.4 & 00:11\\
  2000-03-02T08:54 & 776 & 0.8 & 233 & 62 & 2000-03-02T08:20 & -11 & 70 & X1.1 & 00:11\\
  2000-03-22T19:31 & 478 & -92.0 & 312 & 154 & 2000-03-22T18:34 & 14 & 57 & X1.1 & 00:22\\
   &  &  &  &  & 2000-03-24T07:41 & 16 & 82 & X1.8 & 00:18\\
   &  &  &  &  & 2000-06-06T13:30 & 22 & -10 & X1.1 & 00:16\\
  2000-06-06T15:54 & 1119 & 1.5 & 47 & 360 & 2000-06-06T14:58 & 21 & -9 & X2.3 & 00:42\\
  2000-06-07T16:30 & 842 & 59.8 & 309 & 360 & 2000-06-07T15:34 & 23 & -3 & X1.2 & 00:32\\
  2000-06-18T02:10 & 629 & -1.2 & 318 & 132 & 2000-06-18T01:52 & 23 & 85 & X1.0 & 00:11\\
   &  &  &  &  & 2000-09-30T23:13 & 7 & 91 & X1.2 & 00:15\\
   &  &  &  &  & 2001-04-02T10:04 & 17 & 60 & X1.4 & 00:16\\
  2001-04-02T11:26 & 992 & 3.0 & 278 & 80 & 2001-04-02T10:58 & 15 & 65 & X1.1 & 01:07\\
  2001-10-25T15:26 & 1092 & -1.4 & 175 & 360 & 2001-10-25T14:42 & -16 & 21 & X1.3 & 00:46\\
  2001-12-13T14:54 & 864 & -11.4 & 37 & 360 & 2001-12-13T14:20 & 16 & -9 & X6.2 & 00:15\\
  2002-07-03T02:54 & 265 & -9.9 & 274 & 73 & 2002-07-03T02:08 & -20 & 51 & X1.5 & 00:08\\

\bottomrule
  & & & & & & & & & continued ...\\

\end{tabular}
\end{table}
\end{center}
\end{landscape}

\newpage
\begin{landscape}
\begin{center}
\begin{table}[h!]
\begin{tabular}
{l 
>{\centering\arraybackslash}m{1.2cm} 
>{\centering\arraybackslash}m{1.2cm} 
>{\centering\arraybackslash}m{1.2cm} 
>{\centering\arraybackslash}m{1.2cm} 
c c c c 
>{\centering\arraybackslash}m{2cm} }

\addlinespace
\toprule

   \multicolumn{5}{c}{CME parameters} & 
   \multicolumn{5}{c}{Flare parameters} \\
  First reported &
  \makecell{V \\ (km s\(^{-1}\))} &
  \makecell{Accel. \\ (km s\(^{-2}\))} &
  Mpa (degrees) &
  Width (degrees) &
  Time start &
  Lat &
  Lon &
  Class &
  \makecell{Duration \\ (hrs: mins)}  \\
\midrule

  2002-07-15T20:30 & 1151 & -25.6 & 35 & 360 & 2002-07-15T19:59 & 19 & 1 & X3.0 & 00:15\\
  2002-07-18T08:06 & 1099 & -30.2 & 354 & 360 & 2002-07-18T07:24 & 19 & 30 & X1.8 & 00:25\\
  2002-08-03T19:31 & 1150 & -18.8 & 272 & 138 & 2002-08-03T18:59 & -16 & 76 & X1.0 & 00:12\\
  2002-08-21T06:06 & 268 & -19.6 & 266 & 66 & 2002-08-21T05:28 & -12 & 51 & X1.0 & 00:08\\
  2002-10-31T17:06 & 1061 & -33.4 & 96 & 43 & 2002-10-31T16:47 & 28 & 87 & X1.2 & 00:08\\
  2003-03-17T19:54 & 1020 & -5.5 & 264 & 96 & 2003-03-17T18:50 & -14 & 39 & X1.5 & 00:26\\
  2003-03-18T12:30 & 1601 & -13.3 & 266 & 209 & 2003-03-18T11:51 & -15 & 46 & X1.5 & 00:29\\
  2003-05-27T23:50 & 964 & -9.6 & 67 & 360 & 2003-05-27T22:56 & -7 & 17 & X1.3 & 00:17\\
   &  &  &  &  & 2003-06-09T21:31 & 12 & 34 & X1.7 & 00:12\\
  LASCO data gap &  &  &  &  & 2003-06-10T23:19 & 10 & 40 & X1.3 & 00:53\\
  LASCO data gap &  &  &  &  & 2003-06-11T20:01 & 14 & 57 & X1.6 & 00:26\\
   &  &  &  &  & 2004-02-26T01:50 & 14 & 14 & X1.1 & 00:20\\
   &  &  &  &  & 2004-08-13T18:07 & -13 & 23 & X1.0 & 00:08\\
  2004-08-18T17:54 & 602 & 0.5 & 250 & 120 & 2004-08-18T17:29 & -14 & 90 & X1.8 & 00:25\\
  2004-10-30T12:30 & 427 & 12.4 & 269 & 360 & 2004-10-30T11:38 & 13 & 25 & X1.2 & 00:12\\
   &  &  &  &  & 2005-01-15T00:22 & 14 & -8 & X1.2 & 00:40\\
  2005-09-13T20:00 & 1866 & 11.5 & 149 & 360 & 2005-09-13T19:19 & -11 & -3 & X1.5 & 01:38\\
   &  &  &  &  & 2005-09-15T08:30 & -11 & 24 & X1.1 & 00:16\\
  2011-02-15T02:24 & 669 & -18.3 & 189 & 360 & 2011-02-15T01:44 & -20 & 10 & X2.2 & 00:22\\
   &  &  &  &  & 2011-03-09T23:13 & 8 & 9 & X1.5 & 00:16\\
\bottomrule

\end{tabular}
\end{table}
\end{center}
\end{landscape}

\newpage
\begin{center}
\section{Algorithm 2: False alarms for time range 2}
\label{app:alg2falist80}
\begin{table}[h!]
\caption{List of X-class flares between 1 April 1980 and 31 December 1995 which were false alarms. Column 1 gives the start time of the flare, column 2 its heliographic latitude, and column 3 its heliographic longitude. Column 4 is the class of the flare and column 5 its duration.}  
\begin{tabular}
{l 
c c c 
>{\centering\arraybackslash}m{2cm} }

\addlinespace
\toprule

   \multicolumn{5}{c}{Flare parameters} \\

  Time start &
  Lat &
  Lon &
  Class &
  \makecell{Duration \\ (hrs: mins)}  \\
\midrule

  1980-05-21T20:51 & -14 & 15 & X1.4 & 00:35\\
  1980-05-28T19:24 & -18 & 33 & X1.1 & 01:29\\
  1980-06-04T22:57 & -14 & 69 & X2.2 & 00:17\\
  1980-06-21T01:17 & 20 & 90 & X2.6 & 00:43\\
  1980-07-01T16:22 & -12 & 38 & X2.5 & 00:49\\
  1980-10-14T05:42 & -9 & 7 & X3.3 & 01:52\\
  1980-10-25T09:42 & 19 & 59 & X3.9 & 00:28\\
  1980-11-07T01:56 & 7 & 11 & X2.7 & 01:19\\
  1980-11-08T13:33 & 8 & 28 & X3.3 & 02:05\\
  1980-11-12T04:46 & 10 & 72 & X2.5 & 00:06\\
  1980-11-15T15:40 & -12 & 83 & X1.9 & 01:51\\
  1981-02-17T18:12 & 20 & 20 & X1 & 05:30\\
  1981-02-20T06:40 & 19 & 49 & X2.4 & 01:07\\
  1981-03-25T20:39 & 9 & 89 & X2.2 & 00:44\\
  1981-04-02T11:03 & -43 & 68 & X2.2 & 00:25\\
  1981-07-19T05:32 & -37 & 56 & X2.7 & 01:05\\
  1981-07-26T07:57 & -14 & 18 & X1 & 00:35\\
  1981-07-27T17:24 & -13 & -11 & X1.5 & 01:24\\
  1981-08-12T06:24 & -10 & 28 & X2.6 & 00:56\\
  1981-09-15T21:13 & 10 & 78 & X2.3 & 00:15\\
  1982-02-07T12:50 & -14 & 72 & X1 & 01:21\\
  1982-02-08T12:50 & -13 & 88 & X1.4 & 00:29\\
  1982-02-09T03:57 & -13 & 90 & X1.2 & 00:26\\
  1982-03-30T05:22 & 13 & 11 & X2.8 & 03:04\\
  1982-06-26T00:42 & 16 & 5 & X1.9 & 01:26\\
  1982-06-26T19:09 & 15 & 73 & X2.1 & 01:04\\
  1982-07-17T10:28 & 14 & 32 & X3.2 & 00:53\\
  1982-12-22T08:26 & -9 & 82 & X2.4 & 00:31\\
  1982-12-29T06:43 & -13 & 12 & X1.9 & 00:34\\
  1983-06-06T13:31 & -11 & 15 & X1.4 & 02:01\\
  1988-06-23T08:56 & -19 & 34 & X1.6 & 01:07\\
  1988-06-24T04:18 & -18 & 45 & X1.3 & 02:43\\
  1988-06-24T16:03 & -17 & 52 & X2.4 & 00:51\\
  1988-10-03T14:53 & -27 & 16 & X3.2 & 00:49\\
  1988-10-03T23:22 & -27 & 20 & X1.1 & 00:57\\
  1988-12-30T17:25 & -19 & 30 & X1.4 & 02:23\\  
  
\bottomrule
& & & & continued ... \\
\end{tabular}
\end{table}
\end{center}

\begin{center}
\begin{table}
\begin{tabular}
{l 
c c c 
>{\centering\arraybackslash}m{2cm} }
\addlinespace
\toprule
   \multicolumn{5}{c}{Flare parameters} \\
  Time start &
  Lat &
  Lon &
  Class &
  \makecell{Duration \\ (hrs: mins)}  \\
\midrule

  1989-01-13T08:29 & -31 & 5 & X2.3 & 02:16\\
  1989-01-14T02:54 & -32 & 10 & X2.1 & 02:25\\
  1989-01-14T21:45 & -29 & 26 & X1.1 & 01:24\\
  1989-01-18T07:02 & -30 & 65 & X1.4 & 00:11\\
  1989-01-27T19:08 & -19 & -17 & X1.1 & 01:38\\
  1989-03-14T16:46 & 33 & 21 & X1.1 & 05:02\\
  1989-03-16T15:24 & 36 & 47 & X3.6 & 01:21\\
  1989-03-16T20:35 & 29 & 60 & X1.4 & 00:56\\
  1989-05-05T07:23 & 30 & -1 & X2.4 & 03:12\\
  1989-06-15T18:13 & -21 & -8 & X4.1 & 02:28\\
  1989-06-16T04:19 & -17 & -3 & X3 & 00:26\\
  1989-09-03T14:28 & -18 & -16 & X1.2 & 00:32\\
  1989-09-04T08:57 & -18 & -19 & X1.1 & 00:49\\
  1989-09-09T19:28 & -15 & 67 & X1.3 & 00:28\\
  1989-11-12T06:21 & 13 & 39 & X1.5 & 00:46\\
  1989-11-19T06:19 & -24 & 25 & X1.1 & 00:23\\
  1989-11-20T21:25 & -27 & 43 & X1 & 00:36\\
  1989-11-21T13:32 & -26 & 53 & X4 & 00:59\\
  1989-11-25T22:55 & 30 & -5 & X1 & 03:40\\
  1989-12-30T04:09 & -19 & -9 & X1 & 01:05\\
  1989-12-31T09:32 & -25 & 51 & X2.8 & 00:45\\
  1991-01-30T08:49 & -8 & 34 & X1 & 01:36\\
  1991-01-31T01:58 & -17 & 35 & X1.3 & 03:21\\
  1991-03-16T00:47 & -9 & -9 & X1.8 & 00:22\\
  1991-03-17T20:54 & -10 & 13 & X1 & 02:11\\
  1991-03-29T06:42 & -28 & 60 & X2.4 & 00:52\\
  1991-03-31T19:11 & -22 & 88 & X1 & 00:08\\
  1991-04-20T08:27 & 8 & 50 & X1 & 02:57\\
  1991-05-18T05:06 & 32 & 85 & X2.8 & 02:42\\
  1991-07-31T00:46 & -17 & -11 & X2.3 & 01:29\\
  1991-08-02T03:07 & 25 & -15 & X1.5 & 00:52\\
  1991-09-07T19:11 & -11 & 50 & X3.3 & 01:10\\
  1991-09-08T09:06 & -13 & 58 & X1 & 00:43\\
  1991-10-26T18:53 & -9 & -20 & X1.7 & 04:32\\
  1991-10-27T02:06 & -11 & -20 & X1.9 & 00:51\\
  1991-10-27T05:38 & -13 & -15 & X6.1 & 01:20\\
  1991-11-09T15:32 & -16 & 57 & X1.1 & 01:37\\
  1991-11-15T22:34 & -13 & 19 & X1.5 & 00:43\\
  1991-12-24T10:13 & -17 & -14 & X1.4 & 01:20\\
  1992-01-26T15:23 & -16 & 66 & X1 & 01:02\\
  1992-02-16T12:32 & -13 & 17 & X1.4 & 01:09\\
  1992-02-27T09:22 & 6 & 2 & X3.3 & 03:41\\
  1992-09-06T18:42 & -11 & 41 & X1.7 & 02:09\\
  1992-09-06T20:50 & -11 & 46 & X1.3 & 00:26\\

\bottomrule
\end{tabular}
\end{table}
\end{center}

\end{article}
\end{document}